\documentclass[preprint2]{aastex631}

\usepackage{outlines}

\newcommand{\msec}{m~s$^{-1}$}
\newcommand{\cmsec}{cm~s$^{-1}$}

\newcommand{\heliumAA}{He~10830~\AA}
\newcommand{\helium}{He-10830}
\newcommand{\calcium}{Ca~\textsc{i}~10842~\AA}
\newcommand{\lya}{Lyman-$\alpha$}
\newcommand{\ewhe}{EW[He]}

\newcommand{\oot}{out-of-transit}
\newcommand{\Oot}{Out-of-transit}
\newcommand{\ewdiffspectrum}{``EW difference power spectrum"}
\newcommand{\ewdiffspectra}{``EW difference power spectra"}

\definecolor{myred}{HTML}{BF3465}

\graphicspath{{./}{figures/}}

 \received{June 3, 2022}
 \revised{October 17, 2023}
 \accepted{November 7, 2023}

\submitjournal{AJ}

\shorttitle{nir helium spectroscopy of young stars}
\shortauthors{Krolikowski et al.}
\graphicspath{{./}{figures/}}

\begin{document}


\title{The Strength and Variability of the Helium 10830~\AA\ Triplet in Young Stars, with Implications for Exosphere Detection}

\correspondingauthor{Daniel M. Krolikowski}
\email{krolikowski@arizona.edu}

\author[0000-0001-9626-0613]{Daniel M. Krolikowski}
\affiliation{Department of Astronomy, University of Texas at Austin, 2515 Speedway, Stop C1400, Austin, TX 78751, USA}
\affiliation{Steward Observatory, The University of Arizona, 933 N. Cherry Ave, Tucson, AZ 85721, USA}

\author[0000-0001-9811-568X]{Adam L. Kraus}
\author[0000-0003-2053-0749]{Benjamin M. Tofflemire}
\author[0000-0002-4404-0456]{Caroline V. Morley}
\affiliation{Department of Astronomy, University of Texas at Austin, 2515 Speedway, Stop C1400, Austin, TX 78751, USA}

\author[0000-0003-3654-1602]{Andrew W. Mann}
\affiliation{Department of Physics and Astronomy, The University of North Carolina at Chapel Hill, Chapel Hill, NC 27599, USA}

\author[0000-0001-7246-5438]{Andrew Vanderburg}
\affiliation{Department of Physics and Kavli Institute for Astrophysics and Space Research, Massachusetts Institute of Technology, Cambridge, MA 02139, USA}

\begin{abstract}
Young exoplanets trace planetary evolution, particularly the atmospheric mass loss that is most dynamic in youth. However, the high activity level of young stars can mask or mimic the spectroscopic signals of atmospheric mass loss. This includes the activity-sensitive He~10830~\AA\ triplet, which is an increasingly important exospheric probe. To characterize the He-10830 triplet at young ages, we present time-series NIR spectra for young transiting planet hosts taken with the Habitable-zone Planet Finder. The He-10830 absorption strength is similar across our sample, except at the fastest and slowest rotation, indicating that young chromospheres are dense and populate metastable helium via collisions. Photoionization and recombination by coronal radiation only dominates metastable helium population at the active and inactive extremes. Volatile stellar activity, such as flares and changing surface features, drives variability in the He-10830 triplet. Variability is largest at the youngest ages before decreasing to $\lesssim5-10$~m\AA\ (or 3\%) at ages above 300 Myr, with 6 of 8 stars in this age range agreeing with no intrinsic variability. He-10830 triplet variability is smallest and age-independent at the shortest timescales. Intrinsic stellar variability should not preclude detection of young exospheres, except at the youngest ages. We recommend out-of-transit comparison observations taken directly surrounding transit and observation of multiple transits to minimize activity's effect. Regardless, caution is necessary when interpreting transit observations in the context of stellar activity, as many scenarios can lead to enhanced stellar variability even on timescales of an hour.
\end{abstract}


\section{Introduction}\label{sec:introduction}

Planetary evolution within the first billion years, when stellar activity and planetary orbits are most dynamic, establishes a planet's ultimate atmospheric properties and habitability potential. A planet's birth location in the disk determines the initial mass and composition of atmosphere that can be accreted \citep{oberg2011,ikoma2012}, and its final orbital configuration after any migration sets the planet's radiative environment and atmospheric retention \citep{owen2013,shields2016}. A majority of super-Earth to sub-Neptune sized planets form with a massive H/He envelope that precludes habitability \citep{wolfgang2015,owen2016}. Losing this massive, gaseous envelope is necessary to produce habitable conditions, such as a favorable surface environment and the creation of a secondary atmosphere \citep{pierrehumbert2011,lammer2014}.

High-energy irradiation can sculpt a massive H/He envelope \citep{lammer2003,kubyshkina2018}, leading to photoevaporative mass loss that will alter the atmosphere's size and composition. This mass loss carves the gap in planet occurrence at a radius of $\sim1.7 R_{\oplus}$, separating smaller rocky super-Earths from larger sub-Neptunes with gaseous envelopes \citep{fulton2017}. Photoevaporation is the leading theory for the cause of this atmospheric evolution \citep{owen2013,lopez2013,jin2014}, but mass loss could also be driven by heat from the contracting core or impacts \citep{liu2015,ginzburg2018}.
The exact mechanism sculpting the atmosphere and the timescale over which it acts are crucial inputs for models predicting the habitability of exoplanets \citep{johnstone2015,owen2016}.


One way to distinguish different atmospheric evolutionary pathways is to directly detect ongoing mass loss, and correlate the mass-loss rate with system characteristics, such as the host star's spectrum and planet's orbit. Exoplanet transit spectroscopy can detect excess absorption from gas in the evaporating atmosphere, also known as an exosphere. The first exosphere detections were of escaping hydrogen gas using the \lya\ spectral line \citep[e.g.][]{vidalmadjar2003,lecavelierdesetangs2010,ehrenreich2012,bourrier2018}. Unfortunately, there are difficulties in observing \lya\ exospheres because the line is heavily absorbed by the interstellar medium, is contaminated by geocoronal emission, and requires expensive space-based observations \citep{neff1986,dring1997,ehrenreich2011}. Therefore, \lya\ exospheres are only detectable around the very nearest planets.

Fortunately, there is an increasingly popular probe of atmospheric mass loss that is free of \lya's complications: the helium~10830~\AA\ triplet feature \citep{seager2000,oklopcic2018}, which has been used to detect many exospheres over the last five years \citep[e.g.][]{spake2018,allart2018,salz2018,ninan2020,czesla2022,orellmiquel2022}. This feature is formed from the transition between the two lowest energy He \textsc{i} triplet states (2$^3$S to 2$^3$P). Importantly, the 2$^3$S state is radiatively decoupled from the ground state and thus metastable; a He atom will remain in this state long enough to absorb incoming photons and make the \heliumAA\ transition. Ground-based high resolution NIR spectrographs can access this line, enabling survey observations to map atmospheric mass loss across system demographics.

It is crucial to find young exospheres to reveal the mechanisms driving atmospheric mass loss, because the most significant mass loss is expected within the first Gyr when the stellar high-energy radiative output is highest \citep{micela1985,preibisch2005,jackson2012}. While a subset of the known young transiting planets have been targeted in the search for helium exospheres \citep{hirano2020,gaidos2020a,gaidos2020b,gaidos2021,vissapragada2021,gaidos2022,zhang2022b}, only one robust detection has been made \citep[HD 73583 b;][]{zhang2022a}. This is not surprising, as young stars have high levels of magnetic activity that introduce chromatic, temporally coherent noise across the stellar spectrum \citep{desort2007,reiners2012} that can mask or mimic a planetary signal in transmission spectra \citep{rackham2018,boldt2020}. The stellar \helium\ triplet is chromospheric, and thus is sensitive to stellar activity. This can lead to confusion between a stellar or planetary nature for changes in the \helium\ triplet absorption. The stellar \helium\ triplet of FGKM dwarfs is intrinsically variable \citep{zirin1975,zirin1976,sanzforcada2008,fuhrmeister2020}, but the connection between this variability and activity is unexplored, particularly at young ages. This complicates the interpretation of changes in the \helium\ triplet absorption strength for transit spectra of young planets.

In the stellar chromosphere, there are two main pathways to populate the metastable state: photoionization and recombination (PR), and collisional excitation (CE). In the PR mechanism, a He atom is photoionized (by photons with $\lambda \le 504$~\AA) and then recombines to populate the metastable state. This process is driven by high energy coronal radiation incident upon He atoms in the upper chromosphere \citep{goldberg1939,zirin1975}. The metastable state can be directly populated through CE from the ground state \citep{hartmann1979,wolff1984}, although this requires a dense enough chromosphere for collisions to dominate over the PR mechanism \citep{sanzforcada2008}. In exospheres, the PR mechanism driven by high energy instellation would form the feature because the exosphere density is too low for CE. An exosphere's \helium\ triplet signal is thus dependent on its host star's high energy output, particularly the intensity and shape of the stellar UV spectrum \citep{oklopcic2018,oklopcic2019,poppenhaeger2022}.

Once He atoms populate the metastable state, they can absorb photospheric NIR continuum to excite into the 2$^3$P state and produce the \helium\ triplet absorption feature. Due to splitting of the 2$^3$P state, the feature consists of three lines: two blended lines (rest vacuum wavelengths of 10833.217~\AA\ and 10833.306~\AA), and a weaker resolved component with a rest vacuum wavelength of 10832.057~\AA\ \citep{martin1960}. The feature forms higher in the chromosphere than other traditional spectral chromospheric activity indicators \citep[e.g. Ca \textsc{ii} H and K, H-$\alpha$;][]{dupree1992,avrett1998}, and its formation depends on the structure of the chromosphere, transition region, and corona \citep{avrett1994,andretta1997}.

\helium\ triplet absorption on the Sun is stronger in active regions with higher X-ray emission \citep{andretta1997,mauas2005,andretta2017} and \helium\ triplet equivalent width correlates with X-ray luminosity for inactive FGK dwarfs \citep{zarro1986}, both implying that the PR mechanism is dominant for inactive solar-type stars. This relation does not hold for active FGK dwarfs because their chromospheres are dense enough for CE to dominate \citep{sanzforcada2008}. There is no detectable absorption for SpT $\geq$ M5, due to emission fill-in and reduced line excitation from the PR mechanism \citep{fuhrmeister2019}. 

Volatility of the stellar activity level, from secular changes in the magnetic field and short-term evolution of active events (e.g. surface features, winds, mass loss, or flaring), can cause variability in the \helium\ triplet. This has been observed in FGK dwarfs and giants \citep[although with sparse, telluric-contaminated observations;][]{zirin1976,obrien1986,katsova1998}, and more recently in M-dwarfs \citep[with CARMENES;][]{fuhrmeister2020}. However, the amplitude and timescale of variability remains uncharacterized for young stars with high activity levels.

To better understand the \helium\ triplet in young stars, and the potential for a stellar signal to mask or mimic an exosphere, its absorption strength and variability must be characterized spanning a range of stellar parameters. We have gathered a data set that is well suited for this task: time series NIR spectroscopy of young transiting planet hosts from the Habitable-zone Planet Finder. With these data, we study the \helium\ triplet across age, spectral type, and activity level to explore the effect of stellar activity on the \helium\ triplet and assess the feasibility of detecting young helium exospheres.

\begin{deluxetable*}{ccccccccccc}
\tablecaption{HPF young transiting planet host star sample\label{tab:objects}}
\tablehead{
\colhead{Object} & \colhead{RA} & \colhead{Dec} & \colhead{$J$} & \colhead{Membership} & \colhead{Age\tablenotemark{a}} & \colhead{$T_{\rm eff}$} &
\colhead{$P_{\rm rot}$} & \colhead{$N_{\rm ep}$\tablenotemark{b}} & \colhead{Plot Marker\tablenotemark{c}} & \colhead{Ref\tablenotemark{d}}\\
\colhead{} & \colhead{h m s} & \colhead{d m s} & \colhead{mag} & \colhead{} & \colhead{Myr} & \colhead{K} & \colhead{d} & \colhead{} & \colhead{} & \colhead{}
}
\startdata
\multicolumn{11}{c}{Objects included in helium analysis}\\
\tableline
V1298 Tau & 04 05 19.6 & +20 09 25.6 & 8.7 & Taurus & $28$ & 4970 & 2.851 & 42 & \textcolor[HTML]{BF3465}{$\blacksquare$} & 11,12 \\
K2-284 & 05 16 33.8 & +20 15 18.4 & 10.9 & Field & $120$ & 4140 & 8.88 & 13 & \textcolor[HTML]{731683}{$\blacksquare$} & 10 \\
TOI 2048 & 15 51 41.8 & +52 18 22.7 & 9.9 & Group X & $300$ & 5185 & 7.97 & 13 & \LARGE \textcolor[HTML]{dfa5e5}{$\bullet$} & 14 \\
HD 63433 & 07 49 55.1 & +27 21 47.5 & 5.6 & Ursa Major & $414$ & 5640 & 6.45 & 13 & \LARGE \textcolor[HTML]{1C6CCC}{$\bullet$} & 13 \\
HD 283869 & 04 47 41.8 & +26 09 00.8 & 8.4 & Hyades & $700$ & 4655 & 37 & 15 & \textcolor[HTML]{DFA5E5}{$\blacksquare$} & 8 \\
K2-136 & 04 29 38.9 & +22 52 57.8 & 9.1 & Hyades & $700$ & 4499 & 15 & 36 & \textcolor[HTML]{1C6CCC}{$\blacksquare$} & 5,6,7 \\
K2-100 & 08 38 24.30 & +20 06 21.83 & 9.5 & Praesepe & $700$ & 6120 & 4.3 & 29 & \textcolor[HTML]{db6d1b}{$\blacksquare$} & 4 \\
K2-101 & 08 41 22.58 & +18 56 01.95 & 11.2 & Praesepe & $700$ & 4819 & 10.6 & 3 & \LARGE \textcolor[HTML]{db6d1b}{$\bullet$} & 4 \\
K2-102 & 08 40 13.45 & +19 46 43.71 & 11.3 & Praesepe & $700$ & 4695 & 11.5 & 5 & \LARGE \textcolor[HTML]{bf3465}{$\bullet$} & 4 \\
K2-77 & 03 40 54.8 & +12 34 21.4 & 10.4 & Field & 850 & 4970 & 19.8 & 13 & \LARGE \textcolor[HTML]{731683}{$\bullet$} & 3 \\
\tableline
\multicolumn{11}{c}{Objects not included in helium analysis}\\
\tableline
K2-25 & 04 13 05.6 & +15 14 52.0 & 11.3 & Hyades & $700$ & 3207 & 1.88 & 31 & -- & 1,2 \\
K2-103 & 08 41 38.49 & +17 38 24.08 & 12.3 & Praesepe & $700$ & 3880 & 14.6 & 4 & -- & 4 \\
K2-104 & 08 38 32.84 & +19 46 25.59 & 12.9 & Praesepe & $700$ & 3660 & 9.3 & 4 & -- & 4 \\
K2-264 & 08 45 26.05 & +19 41 54.46 & 13.1 & Praesepe & $700$ & 3580 & 22.8 & 6 & -- & 9 \\
EPIC 211901114 & 08 41 35.69 & +18 44 34.98 & 13.2 & Praesepe & $700$ & 3440 & 8.6 & 3 & -- & 4 \\
K2-95 & 08 37 27.06 & +18 58 36.02 & 13.3 & Praesepe & $700$ & 3410 & 23.9 & 4 & -- & 4 \\
\enddata
\tablenotetext{a}{We adopt the same age for Hyades and Praesepe, 700 Myr, which is roughly the average of previous age determinations \citep{brandt2015,martin2018,gossage2018}. The age for V1298 Tau comes from \citet{johnson2022}, and all other ages are from the planet discovery papers.}
\tablenotetext{b}{Number of in-hand and usable visits for each target through 10/18/2021.}
\tablenotetext{c}{Marker and color used for each object's plotted point in Figures~\ref{fig:he_val_teff}, \ref{fig:he_val_prot}, \ref{fig:he_ew_scatter}, \ref{fig:he_ew_excess_scatter_prot}, \ref{fig:intra_visit_scatter}}
\tablenotetext{d}{Planet discovery references: (1) \citet{mann2016a}, (2) \citet{david2016a}, (3) \citet{gaidos2017}, (4) \citet{mann2017}, (5) \citet{mann2018},
(6) \citet{ciardi2018}, (7) \citet{livingston2018}, (8) \citet{vanderburg2018}, (9) \citet{rizzuto2018}, (10) \citet{david2018}, (11) \citet{david2019a}, (12) \citet{david2019b},
(13) \citet{mann2020}, (14) \citet{newton2022}}
\end{deluxetable*}

\section{Description of stellar sample and HPF observations}\label{sec:observations}

We present observations of 10 young ($\tau < 1$~Gyr) transiting planet hosts, comprising all such known systems as of July 2020 that are observable with the HET at McDonald Observatory ($-10\arcdeg < \delta < 72\arcdeg$). We obtained these observations for a survey to search for outer giant planets in these systems, which will be described in a forthcoming paper. These 10 stars are a subset of a larger 16 target sample used in the RV planet search survey. The 6 stars that are excluded are faint enough that it is difficult to measure precise and reliable spectral line equivalent widths. This removes all M-dwarfs from our sample, so our study here concerns only FGK dwarfs. The stars in this paper's sample span: $\sim25$~Myr to 1~Gyr in age, late-F to late-K in spectral type, and 1.88~days to 37~days in rotation period. Information for the full sample is listed in Table~\ref{tab:objects}, including the targets that are excluded in this analysis.

We obtained NIR spectra with the Habitable-zone Planet Finder \citep[HPF;][]{mahadevan2012,mahadevan2014} on the 10-m Hobby-Eberly Telescope at McDonald Observatory. HPF is a high-resolution ($R\sim55000$), fiber-fed, stabilized NIR spectrograph covering $z, Y, J$ bands from $8100-12700$~\AA\ \citep{hearty2014,stefansson2016}. HPF has a laser frequency comb (LFC) to achieve extremely high quality wavelength calibration, exhibiting $\sim20$~\cmsec\ calibration precision and $\sim1.5$~\msec\ on sky precision for Barnard's Star \citep{metcalf2019}. HPF has three fibers: a science, sky, and calibration fiber. The sky fiber provides simultaneous observations of nearby sky to subtract sky background (e.g. Moon contamination or airglow), and the calibration fiber can observe the LFC to provide an instantaneous wavelength solution.

Our observational set up is designed for a precision RV planet search. For targets brighter than $J = 10$ we use an exposure time of 5 minutes, which is the maximum allowed exposure time while simultaneously observing the LFC to avoid LFC saturation. This strategy optimizes target S/N with a precise, instantaneous wavelength calibration. The sole exception is for HD 63433, which is bright enough to reach high S/N in only 3 minutes. The 5 minute exposure scheme was used for targets fainter than $J = 10$ in the first year of observations. From August 2019 onward, we use an exposure time of 10 minutes to reach higher S/N for these fainter objects. We bracket these 10 minute exposures with LFC-only observations to improve the wavelength calibration. This method can correct for HPF drift and provides similarly precise wavelength calibration as LFC-simultaneous observations \citep{stefansson2020}.

These observations are made using HET's flexible queue-mode scheduling \citep{shetrone2007}, which allows for observations to be spread throughout the trimester to cover baselines of days to months. Each observation (which we call a ``visit") is composed of three sequential exposures to increase signal and average over high-frequency, oscillation-driven stellar RV variation. The sole planned exception to this is an observation of V1298 Tau that attempted to measure Planet b's transit on UT 2019/10/24, during which we obtained 11 sequential exposures. However, based on an updated prediction for the time of planet b's transits from TESS data \citep{feinstein2022}, this visit did not occur during transit. Occasionally, there were other unplanned exceptions to our three-exposure-per-visit strategy, such as when additional exposures were taken due to poor S/N from passing clouds.

The raw 2D images were reduced to 1D spectra using the HPF team's custom pipeline following the procedures of \citet{ninan2018}, \citet{kaplan2019} and \citet{metcalf2019}. We correct the derived wavelength solution for barycentric motion using barycentric velocities calculated at each exposure time using \texttt{barycorrpy} \citep{kanodia2018}, which is a Python implementation of the formalism from \citet{wright2014}.

We began observations in November of 2018, and the program is still ongoing. All targets have time series spanning at least 9 months, with roughly half of the targets having observations spanning the entire two and a half year program. This unique data set -- namely time series, high resolution NIR spectra of young planet hosts -- enables us to study the strength and intrinsic variability of the stellar \helium\ triplet at young ages. 

\section{measuring spectral line equivalent widths}\label{sec:measurements}

To study the behavior of the \helium\ triplet at young ages, we measure the feature's equivalent width (EW) from the HPF spectral time series for each star. For this analysis, we must first correct the spectra for telluric contamination before measuring the EWs. We describe our procedures for both of these steps in this section.

\subsection{Correcting telluric contamination}\label{sec:tellurics}

The region of interest around the \helium\ triplet contains significant telluric contamination, including sky emission and water absorption. These telluric features can significantly affect the measurement precision of the \helium\ triplet's EW and line profile shape, the latter of which is needed to detect velocity structure and study planetary outflow dynamics. There are drawbacks to both empirical and theoretical telluric modeling, and adequately removing the contamination is a difficult task.

One way to avoid telluric contamination is to observe targets when the strong features do not directly overlap the \helium\ triplet, due to the Doppler shift of the stellar spectral lines from relative motion between the target and the Earth. However, this strategy cannot be used for all targets because it is possible that the barycentric motion of a given object during its peak observability places the telluric features within the \helium\ triplet region. This can be an even more pressing issue for the time-sensitive constraints of transit observations.

This is particularly a problem for the nearby young clusters and associations hosting a majority of the known young transiting planets. We calculated the wavelength of the strong sky emission line near the \helium\ triplet in the rest-frame of a variety of these young stellar associations  (Hyades, Praesepe, Pleiades, Taurus, Ursa Major, and Group X) across the year for observations taken at McDonald Observatory. On average, the emission line is 1~\AA\ from the strong, red component of the \helium\ triplet, ranging from 2.7~\AA\ to directly overlapping the He line. The young clusters along the ecliptic that were accessible by K2 feature the worst contamination. This highlights the importance of TESS's all-sky search which includes many other young stellar associations (including Ursa Major and Group X). It is therefore crucial to adequately correct for this telluric contamination. Below, we describe our methods for removing sky emission lines (Section~\ref{sec:skysubtraction}) and water absorption features (Section~\ref{sec:telluricabsorption}).

\begin{figure}
    \centering
    \includegraphics[width=\columnwidth]{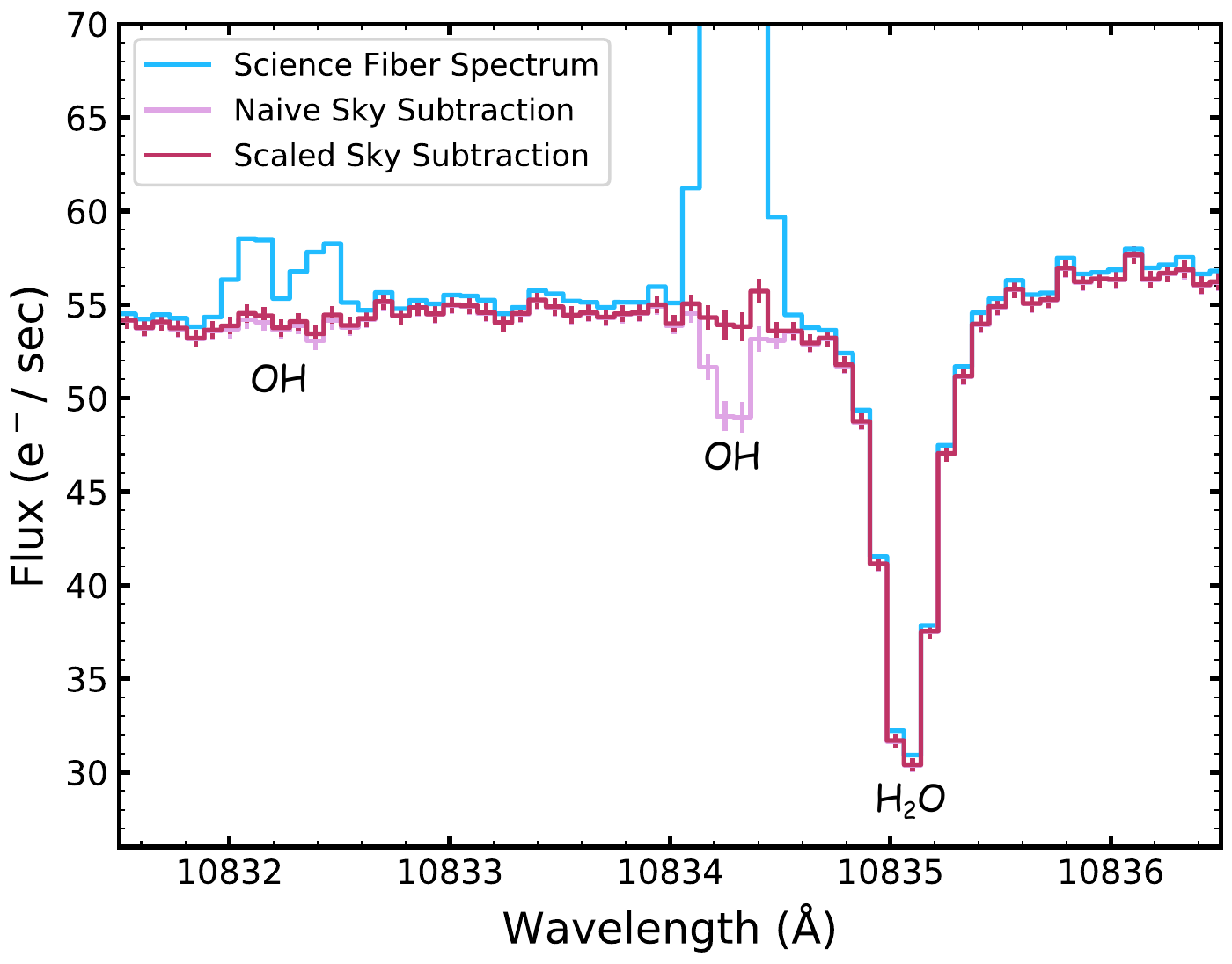}
    \caption{The spectral region around the \helium\ triplet for the A2 star HR 5162. We show an A-star to highlight the emission line over-subtraction as it is devoid of stellar absorption lines. The cyan spectrum is the science fiber flux prior to sky subtraction, which shows the sky emission lines, and the pink spectrum is the sky-subtracted spectrum following naive subtraction. For the strongest sky line, the over-subtraction is roughly 10\% of the nearby continuum, which is significantly larger than the measurement error. The red spectrum shows the sky-subtracted spectrum using the scale factor $\beta$ determined from the sky emission line at 10289.5~\AA. The scaling removes the over-subtraction, and the resultant spectrum agrees with the nearby continuum within errors. Both subtraction methods are adequate and agree for the nearby weak emission doublet.}
    \label{fig:oversub_example}
\end{figure}

\subsubsection{Subtraction of sky emission lines}\label{sec:skysubtraction}

There are three sky emission lines in the \helium\ triplet spectral region: one strong line (which is actually an unresolved doublet) with a rest vacuum wavelength of 10834.2895 \AA, and a weaker resolved doublet at 10832.103 \AA\ and 10832.412 \AA\ \citep{oliva2015}. In principle, the presence of sky emission lines should not be an issue because HPF has a dedicated fiber to simultaneously observe blank sky during every visit. The spectrum observed by this fiber, $f_{\rm sky}$, could then be subtracted from the science fiber spectrum, $f_{\rm sci}$, to produce a telluric emission free spectrum, $f_{\rm skysub}$. However, this naive sky subtraction often results in an over-subtraction of the emission lines in HPF data, introducing residual absorption artifacts in the resultant spectrum. Figure~\ref{fig:oversub_example} shows the result of a naive sky subtraction with HPF data of the A-star HR 5162. It is difficult to assess the amount of over-subtraction in stars of later type due to the multiple stellar lines near the \helium\ triplet, but the lack of stellar features in the A-star spectrum highlights the over-subtraction.

The degree to which the emission lines are over-subtracted depends on the relative flux from the target and the sky, which is demonstrated by the lack of over-subtraction for the weak emission doublet in Figure~\ref{fig:oversub_example}. The over-subtraction residual approaches the measurement uncertainty, and eventually disappears, as the ratio of the target flux to the peak of the sky emission line increases. This means that the brightest targets will have no significant over-subtraction. Therefore, over-subtraction is a greater issue for bright sky emission lines and ``faint" targets. This applies to many of our observations: the emission line at 10834.2895 \AA\ is bright enough, and many of our young transiting planet host sample are faint enough, for the over-subtraction to be significant.

A better method for sky subtraction would involve scaling the sky fiber flux to account for the difference between the science and sky fiber fluxes (such as from throughput differences). Most simply, we can do this by multiplying the entire sky spectrum by a scalar value, which we call $\beta$. The equation describing the sky subtraction is then:

\begin{equation}\label{eq:skysub}
    f_{\rm skysub} = f_{\rm sci} - \beta f_{\rm sky}
\end{equation}

To determine $\beta$ for each observation, we could fit the spectrum around the sky emission line and find the $\beta$ value that results in a spectrum closest to the nearby continuum. However, we cannot use the region around the \helium\ triplet to determine $\beta$ because there is no clear continuum due to the presence of multiple stellar and telluric absorption features.

\begin{figure}
    \centering
    \includegraphics[width=\columnwidth]{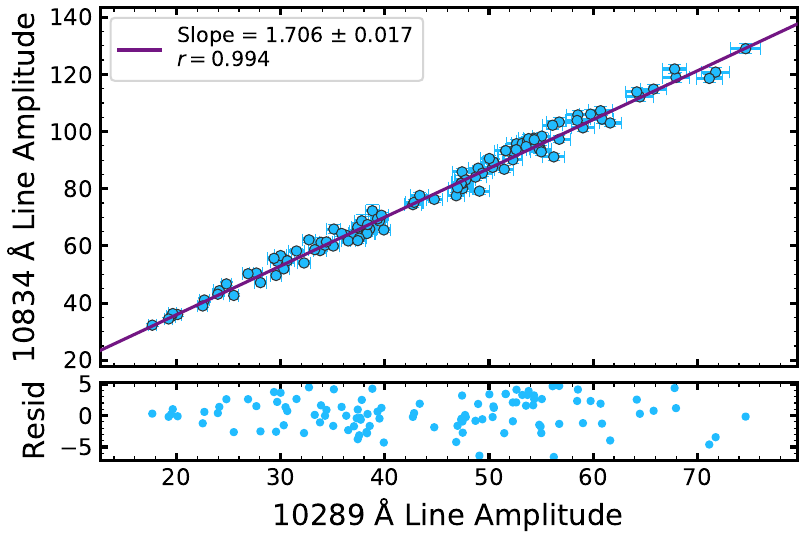}
    \caption{Comparison of the amplitudes for the strong sky emission line near the \helium\ triplet ($\lambda = 10834.3$~\AA) and the calibration sky emission line ($\lambda = 10289.5$~\AA) used to determine the sky subtraction scale factor $\beta$. The amplitudes of the emission lines are well-correlated, and their relationship is linear over an order of magnitude in amplitude.}
    \label{fig:skyline_amp_comp}
\end{figure}

We instead use a different strong sky emission line that is isolated from other spectral features to measure $\beta$. We assume that the OH sky emission lines all behave similarly because their strength should largely be a product of sky conditions, although there may be slight differences depending on the physics of the transitions. We searched the full HPF spectral range for a sky emission line of similar strength to the line at 10834~\AA\ to use as a calibrator line. While there are many strong sky lines throughout the HPF bandpass, a vast majority of them reside in regions with many stellar and telluric absorption features. We were able to identify one high quality candidate calibration sky emission line without any nearby absorption features: an unresolved doublet with a rest vacuum wavelength of 10289.455~\AA. Figure~\ref{fig:skyline_amp_comp} shows a comparison of the two sky emission line amplitudes using the sky fiber spectra from our observations of K2-136. The amplitudes are well-correlated with minimal scatter, and we conclude that this sky emission line is an adequate calibrator with which to determine $\beta$.

To calculate $\beta$, we find the scale factor needed to match the calibration line flux in the sky fiber spectrum to that of the science fiber spectrum (thus accounting for underlying continuum flux). We perform a minimization to find the best fit $\beta$ value, represented as:

\begin{equation}\label{eq:beta}
    \min_{\beta}~{\rm stdev} \left( \frac{f_{\rm sci} - \beta f_{\rm sky}}{f_{\rm cont}} \right)
\end{equation}

\noindent We perform the minimization over a 2.5~\AA\ region around the calibration line between 10288.105~\AA\ and 10290.605~\AA. The sky fiber spectrum is scaled and subtracted from the science fiber spectrum, producing a sky-subtracted residual spectrum. This residual spectrum is fit with a line to obtain a rough estimate of the continuum level in the region, $f_{\rm cont}$. The best fit $\beta$ value is the scale factor such that the standard deviation of the continuum-normalized, scaled sky-subtracted spectrum is minimized. We use the adopted $\beta$ value to scale the entire sky spectrum and subtract it from the science fiber spectrum to produce the sky-free spectrum as described in Equation~\ref{eq:skysub}. 

Figure~\ref{fig:oversub_example} shows the scaled sky subtraction of the A-star HR 5162 in the \helium\ triplet region. The scaling factor works well, and the sky-free spectrum at the location of the strong sky emission line agrees with the nearby continuum value within measurement uncertainty. This demonstrates that the $\beta$ value determined using the calibrator can be applied to the rest of the sky spectrum, and we perform this corrected sky subtraction method for nearly all data presented in this paper. The only exception is for the target HD 63433, which is bright enough for over-subtraction to not be a significant issue. Thus, we naively subtract the sky emission for all spectra of HD 63433.

\begin{figure*}
    \centering
    \includegraphics[width=\textwidth]{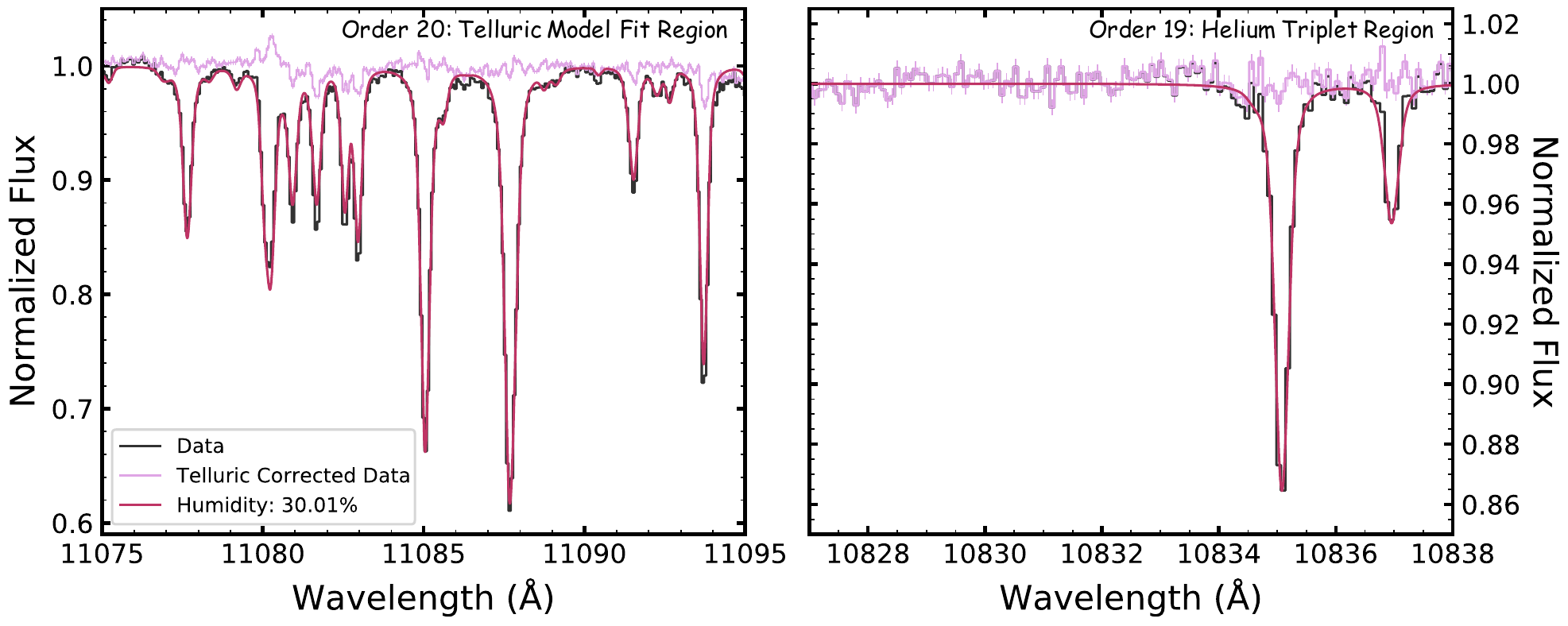}
    \caption{Telluric absorption model fitting for the A-star HR 3711. The black spectrum shows the sky-subtracted, continuum-normalized data, the red spectrum is the best-fit telluric model, and the pink spectrum is the telluric-corrected spectrum. \textit{Left Panel:} The region in spectral order 20 used to find the best-fit model. There are slight residuals likely due to continuum division or theoretical model imperfections. \textit{Right panel:} The \helium\ triplet spectral region. The two strong water lines are well-corrected. The resulting residuals across the range shown is 0.4\%, which is very close to the typical measurement uncertainty of 0.3\%.}
    \label{fig:telluric_fit_example}
\end{figure*}

\subsubsection{Telluric absorption correction}\label{sec:telluricabsorption}

There are also two water absorption features in the \helium\ triplet region that can affect EW measurement: a strong line with a rest vacuum wavelength of 10835.08~\AA\ that can often overlap the \helium\ triplet and a weaker line with a rest vacuum wavelength of 10836.95~\AA. Telluric standard observations are not taken regularly with HPF as the overhead for these observations is impractical for typical programs. Without observed standards, we can not empirically correct for telluric absorption features in our data. Masking the spectra near strong telluric absorption features would exclude important spectral information, especially because the strong water line often directly overlaps the \helium\ triplet profile.

Given these analysis constraints, we decided to generate theoretical telluric absorption models that are custom fit for each individual observation. We use the \texttt{TelFit} python package \citep{telfit}, which can generate telluric absorption models and iteratively fit data for the best parameter values. We only fit for the humidity value because the humidity provided in the observation header is often not close to producing a well-fitting model, and humidity has the largest effect on the telluric model itself. We fix the temperature and pressure values to those recorded in the observation headers. During the fitting process, we fix the spectral resolution to the instrument resolution of 55000.

Using 43 A-star observations from a telluric standard library that has been slowly built since HPF's commissioning, we ran telluric fits on all spectral orders except for those with very strong telluric absorption bands to assess the fit quality across wavelength. We find a wavelength dependence in the best fit humidity value, where the 6 orders blueward of 1~\micron\ produce a consistently lower humidity value than the 5 orders redward of 1~\micron. This is perhaps due to systematic differences in the molecular transition database across wavelength, and is unlikely to be due to the varying presence of stellar spectral features which are largely absent in A-star spectra. We find that the spectral order containing the \helium\ triplet provides humidity values with large scatter relative to the mean humidity value across all orders, indicating that it is not reliable for the fitting process. The typical value found in the \helium\ triplet's order is similar to the subsequent redder order, though, so we use that order for all telluric model fits moving forward. 

For the final telluric absorption model fitting, we use a 20~\AA\ wide region in spectral order 20 covering $11075-11095$~\AA. We use this small region because it has many non-saturated water lines that are separated enough to avoid issues from blending, while having little stellar contamination. By operating on a small spectral segment, we also avoid additional error introduced from the continuum and blaze correction on a full spectral order. With the best fit telluric model parameters from order 20, we then generate a telluric model for the entire HPF bandpass. Figure~\ref{fig:telluric_fit_example} shows an example of the telluric absorption model fit for the A-star HR 3711. Two spectral regions are shown: the region in spectral order 20 where the fit is performed, and the \helium\ triplet region in spectral order 19. The two water lines in the \helium\ triplet region are well-corrected using the model, leaving us confident in our procedure for removing telluric absorption.

\subsection{Measuring the equivalent width of spectral lines in the \helium\ triplet region}\label{sec:measure_ew}

With telluric contamination removed, we can precisely measure the EW of the \helium\ triplet. The spectral region around the \helium\ triplet has other stellar spectral lines which complicates the measurement of the isolated \helium\ triplet EW. The nearby stellar and telluric features also make it more difficult to define the continuum level directly adjacent to the \helium\ triplet. We therefore fit the entire spectral region from 10822~\AA\ to 10845~\AA, including all stellar spectral lines, to provide the best leverage for fitting the continuum level and measuring EWs. This also results in EWs for a handful of other presumably inactive lines that provide useful comparison to the active \helium\ triplet.

\begin{deluxetable}{ccc}
\tablecaption{Spectral lines in the fit region\label{tab:spectral_lines}}
\tablehead{ \colhead{Element} & \colhead{$\lambda$\tablenotemark{a}} & \colhead{Line Profile}\\
\colhead{} & \colhead{(${\rm\AA}$)} & \colhead{}}
\startdata
Cs \textsc{ii}\tablenotemark{b} & 10824.6926& Gaussian\\
Si \textsc{i} & 10830.054 & Lorentzian\\
He \textsc{i} & 10832.057 & Gaussian\\
He \textsc{i} & 10833.2615\tablenotemark{c} & Gaussian\\
Ti \textsc{i} & 10836.38 & Gaussian\\
Na \textsc{i} & 10837.8435\tablenotemark{d} & Gaussian\\
Ca \textsc{i} & 10841.95 & Gaussian\\
\enddata
\tablenotetext{a}{Vacuum line center wavelength taken from the NIST atomic spectral line database \citep{nistatoms}.}
\tablenotetext{b}{This line may be blended with a nearby Cr \textsc{i} line.}
\tablenotetext{c}{The average wavelength of the two blended red components of the helium triplet.}
\tablenotetext{d}{The average wavelength of two close Na \textsc{i} transitions.}
\end{deluxetable}

In addition to the two resolved components of the \helium\ triplet, there are 5 other prominent stellar spectral lines in this wavelength range, although their strength varies across the range of effective temperatures covered in our sample. All of the stellar spectral lines included in fits of this region are listed in Table~\ref{tab:spectral_lines}. The nearby strong silicon line is the closest spectral feature to the \helium\ triplet in wavelength, which it overlaps with its broad wings. The other lines do not directly contaminate the \helium\ triplet, and are relatively weak and narrow. 

Our spectral model has three components: 1) the combined profiles of the 7 spectral lines listed in Table~\ref{tab:spectral_lines}, 2) the telluric model that is separately fit for each individual exposure (as described in Section~\ref{sec:telluricabsorption}), and 3) a 2nd order polynomial continuum. We model the silicon line profile with a Lorentzian due to its wide wings, and the other 6 line profiles with Gaussians. In our data processing pipeline, we use an iterative b-spline to fit the continuum level of the entire \helium\ triplet spectral order. For this subset of the spectral order, we initialize the continuum parameters by fitting a polynomial to the b-spline continuum only for the 23~\AA\ region we study here. The full model is described as:

\begin{equation}\label{eq:spectrum_model}
    f(\lambda)_{\rm model} = C(\lambda)~f(\lambda)_{\rm tell}~\sum_{i=1}^{N_{\rm lines}}{p(\lambda, A_i, \mu_i, s_i)}
\end{equation}

\begin{equation}\label{eq:continuum_equation}
    C(\lambda) = c_0 + c_1\lambda + c_2\lambda^2
\end{equation}

\noindent where $C(\lambda)$ is the continuum polynomial, $f(\lambda)_{\rm tell}$ is the telluric absorption model described in Section~\ref{sec:telluricabsorption}, and the summation represents the combined spectral line profiles. In the summation, $N_{\rm lines}$ is the number of spectral lines included in the fit, $p$ is the profile used to describe each particular spectral line listed in Table~\ref{tab:spectral_lines}, $A$ is the amplitude of the line profile, $\mu$ is the wavelength of the spectral line, and $s$ is the scale factor of the line profile. The scale factor is given by the standard deviation of a Gaussian profile and the scale parameter of a Lorentzian profile. In all, the free parameters in our model are the three continuum polynomial coefficients ($c_0, c_1, c_2$) and the three parameters describing each spectral line included in the model ($A_i, \mu_i, s_i$). We explicitly remind readers that the telluric absorption model is fit independently prior to the procedure described here.

We modify the model described in Equation~\ref{eq:spectrum_model} for two stars in our sample: K2-100 and V1298 Tau. Both of these stars are rapid rotators, and a stellar rotational broadening profile is not a Gaussian or a Lorentzian. Thus, we rotationally broaden the model using literature measurements of their $v\sin i$. For K2-100, we exclude the Cs and Ti spectral lines because they are too weak at the star's effective temperature to be detectable given the rotational broadening. For V1298 Tau, we exclude the Ti line and the blue component of the \helium\ triplet, which are completely blended with nearby lines. We also modify the continuum for V1298 Tau to be a line, because it is so rapidly rotating that a 2nd order polynomial overfits the blended helium and silicon absorption.

\begin{figure}
    \centering
    \includegraphics[width=\columnwidth]{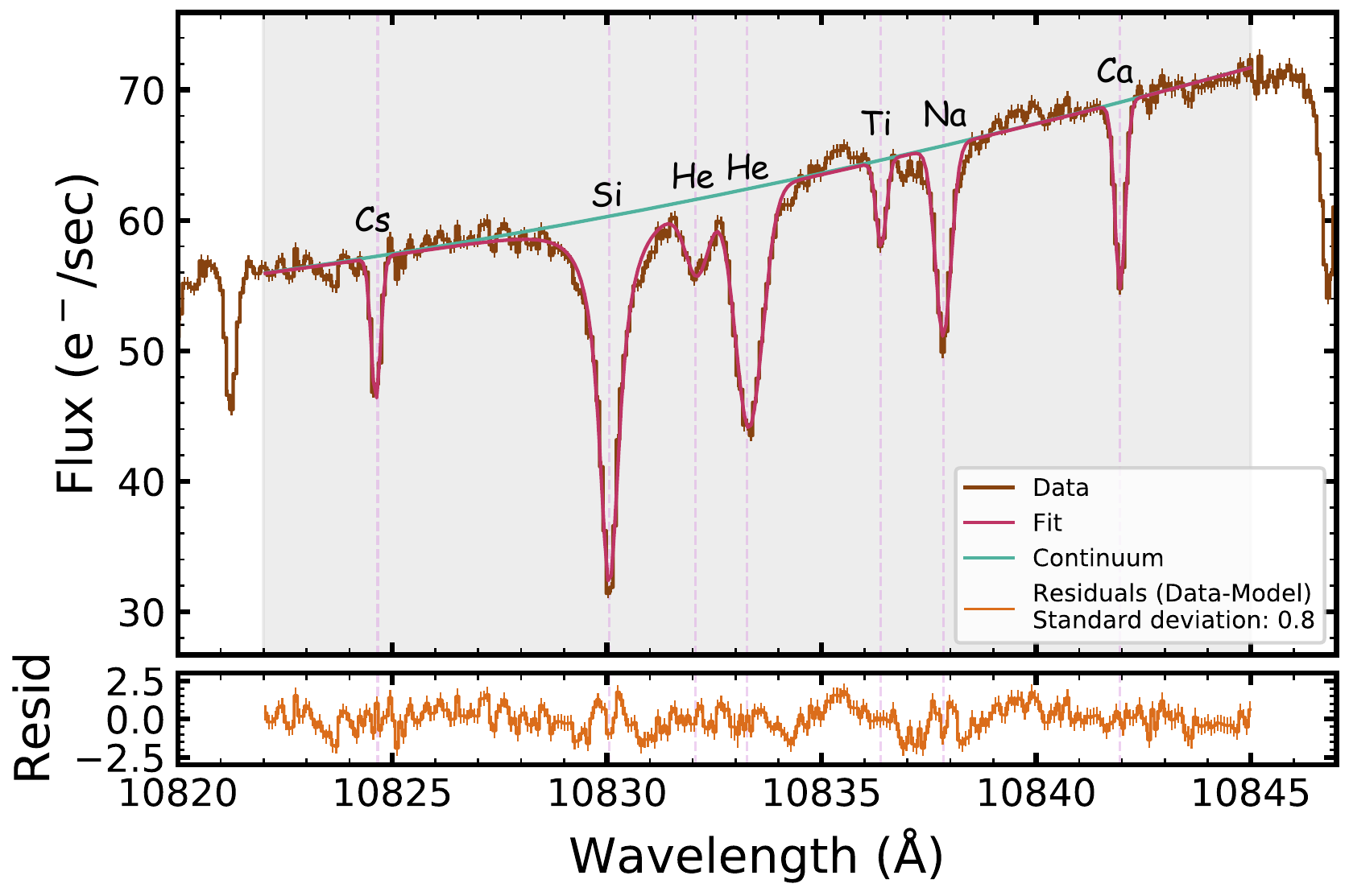}
    \caption{An example result for the spectral fitting of the \helium\ triplet region on an observation of K2-136. The brown spectrum is the telluric-corrected data, the red line is the best fit spectral model, and the teal line is the continuum polynomial. The telluric absorption model has been removed from the data and spectral model to highlight the stellar lines. The bottom panel shows the residuals from subtracting the model from the data. The standard deviation of the residuals is 0.8 e$^-$/s, which corresponds to a percent error on the flux of 1.4\%. This is slightly larger than the typical measurement error of 0.5 e$^-$/s (0.9\%). 
    The continuum level and stellar line profiles are well-matched by the model.}
    \label{fig:spectrum_fit_example}
\end{figure}

We fit the observed flux after scaled sky subtraction, $f_{\rm skysub}$ (Equation~\ref{eq:skysub}), with the model described in Equation~\ref{eq:spectrum_model} using the least squares optimization implemented in \texttt{curve\_fit} from the \texttt{scipy} package \citep{scipy}. We use the implementation of the Gaussian and Lorentzian profiles from \texttt{astropy} \citep{astropy2013,astropy2018}. We include uncertainties on the observed flux which are calculated by adding in quadrature the errors on the science and sky fiber spectra. Figure~\ref{fig:spectrum_fit_example} shows an example fit of an observation of K2-136. The resultant model captures the continuum level and shape, and adequately fits the individual spectral line profiles. 

We compute EWs for all spectral lines included in each particular star's fits by numerically integrating the observed spectra. We first divide the sky-subtracted spectrum by the best fit model, and fit this residual spectrum with a 2nd order polynomial to remove any remaining continuum shape that may bias the numerical EWs. To assess the quality of the spectral fit and secondary continuum correction, we measure the EW of both the corrected and uncorrected residual spectra, which should be 0. The median residual spectrum EW across all observations is $9.2\pm17.9$~m\AA\ without the secondary continuum correction, and $-0.23\pm0.76$~m\AA\ following the correction. The EW for the corrected residual spectrum is much closer to 0 and has significantly less variation across observations, showing that our secondary continuum correction is necessary to capture remaining continuum deviations that could affect the numerical EW integration.

For each individual element, we generate a model that includes all spectral lines \emph{except for that element's feature}; for helium this includes the two Gaussian lines for each resolved component of the triplet (except for V1298 Tau which only fits one helium component). We then compute the residual spectrum using this modified model, which will leave only that element's feature in the data, and numerically integrate to compute the EW. We perform sampling, using draws from the spectral model fit parameters and covariance matrix to calculate 1000 residual spectra for each element. We adopt the median and the median absolute deviation (scaled by 1.4826 to be statistically equivalent to the Gaussian standard deviation) of the EW samples as the value and uncertainty.

By direct numerical integration of the data, the EWs will not be affected by potential mismatch between the actual and modeled line profiles. This also accounts for any asymmetries in the spectral lines included in our fit, such as for the inherent asymmetry of the redder component of the \helium\ triplet. As a comparison check, we also computed the EWs analytically from the fit parameters for each spectral line profile. The numerical and analytical values agree within uncertainties. Given the advantages that the numerical EW calculation has with using the data's line profile, we are confident in adopting the numerical EWs for this paper.

\begin{figure}
    \centering
    \includegraphics[width=\columnwidth]{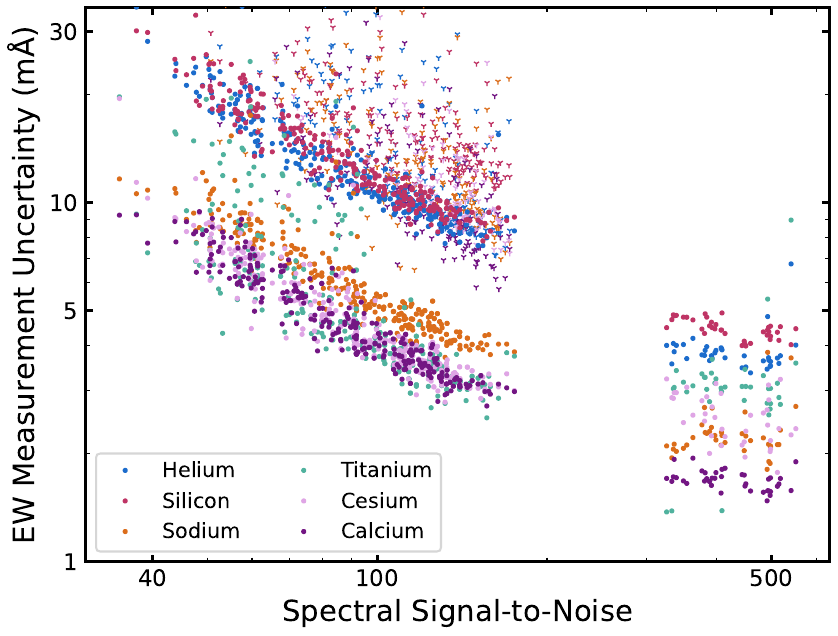}
    \caption{The measurement error of each spectral line for all targets as a function of the median spectrum signal-to-noise in the \helium\ triplet order. The 3-pronged smaller markers are measurements for K2-100 and V1298 Tau, which feature larger error than the rest due to their rapid rotation. The inverse relationship expected is shown, and the uncertainty plateaus at $\la5$~m\AA\ for the \helium\ triplet above S/N $\sim300$.}
    \label{fig:ew_error}
\end{figure}

Figure~\ref{fig:ew_error} shows the EW measurement uncertainty of each spectral line as a function of the median signal-to-noise in the \helium\ triplet spectral order for all targets. The uncertainty decreases with increasing signal-to-noise, as expected. The typical uncertainties are larger for helium and silicon compared to the other lines, at least partially due to the fact that they have significantly larger EWs. The two features are also broader, and the silicon line in particular features strong absorption wings. Regardless, the EW uncertainty is in general precise enough to adequately measure the variability of our time series data, and it plateaus at $\sim5$~m\AA\ for the highest quality spectra. Given the typical He EW value ($\sim325$~m\AA), a single-epoch measurement error of $5$~m\AA\ produces a 1.5~\% error in the EW value. 

For the rest of the paper, we denote the \helium\ triplet EW as \ewhe; for the objects with both resolved components fit, this value is the sum of the individual component EWs. Table~\ref{tab:object_helium_ews} shows all \ewhe\ measurements, and provides an explanation for observations without a calculated \ewhe\ (from low spectral S/N or a failed spectral fit).

\begin{deluxetable}{cccc}
\tablecaption{Helium 10830~\AA\ Triplet EW Measurements\label{tab:object_helium_ews}}
\tablehead{
\colhead{Date (JD)} & \colhead{EW[He]} & \colhead{$\sigma_{\rm EW[He]}$} & \colhead{Flag\tablenotemark{a}}\\
\colhead{} & \colhead{m\AA} & \colhead{m\AA} & \colhead{}
}
\startdata
\tableline
\multicolumn{4}{c}{K2-136}\\
\tableline
2458425.972 & 315.26 & 10.43 & - \\
2458425.976 & 307.7	& 12.42 & - \\
2458425.98 & 326.86	& 12.29 & - \\
\tableline
\multicolumn{4}{c}{V1298 Tau}\\
\tableline
2458548.609	& 409.43 & 10.77 & - \\
2458548.613 & - & - & 1 \\
2458548.618 & 456.2 & 36.62 & - \\
\enddata
\tablenotetext{}{Small sample of EW measurements shown, all measurements are in the machine readable table file.}
\tablenotetext{a}{Flag denoting the reason for missing EW measurement: (1) Low spectrum S/N, (2) failed spectral fit, (3) bad spectral fit.}
\end{deluxetable}


\section{Time series properties of the He triplet}\label{sec:results}

The line profile of the \helium\ triplet probes the conditions in the stellar atmosphere from which the line arises. Different line formation pathways (e.g. the photoionization-recombination mechanism or collisional excitation) will set the dependence of absorption strength on various stellar properties and is indicative of the stellar activity level \citep{zarro1986,sanzforcada2008}. Evolution in the stellar activity can cause variability in the line strength, and the amplitude and timescale of variation can indicate the types of active features that affect the \helium\ triplet \citep{zirin1976,fuhrmeister2020}. With our sample, we have a unique window into the \helium\ triplet in \textit{young} active stars spanning age and spectral type.


\subsection{\helium\ triplet absorption strength}

\begin{figure}
    \centering
    \includegraphics[width=\columnwidth]{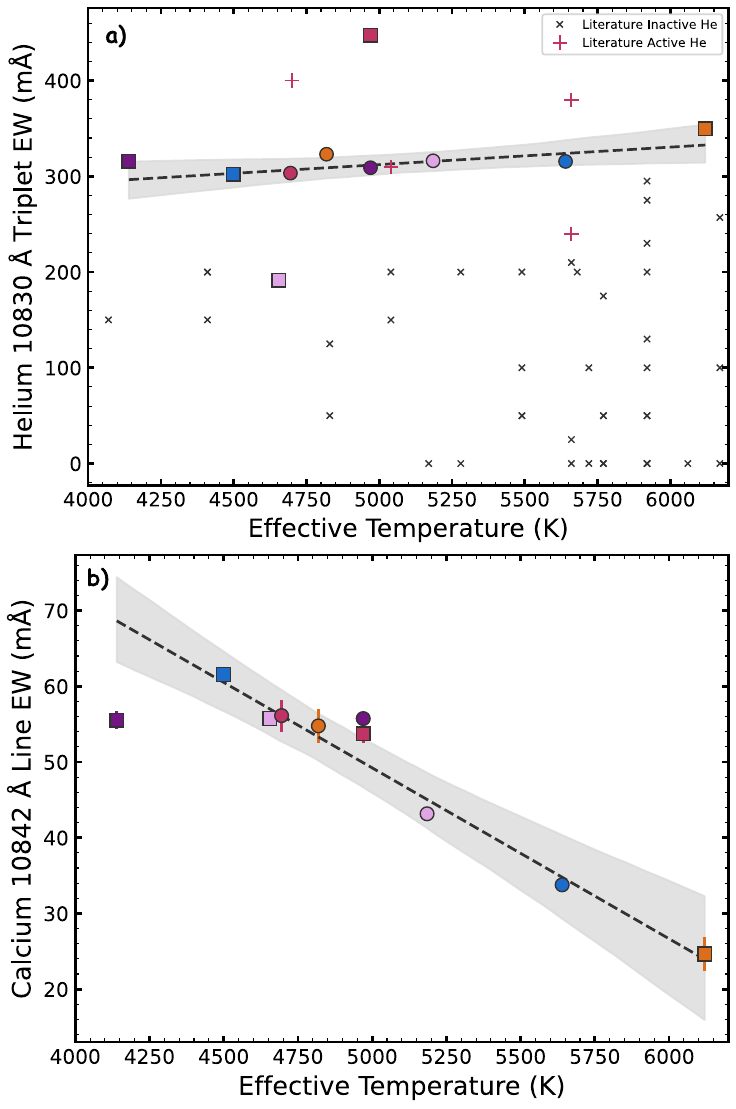}
    \caption{\ewhe\ (top panel) and the \calcium\ line (bottom panel) as a function of effective temperature. Our sample is plotted as large colored circles and squares, and values are taken as the median of the time series measurements. The uncertainties are often smaller than the points. In the top panel, the black x's and red +'s show literature \helium\ triplet EWs for field inactive \citep{zarro1986} and active \citep{sanzforcada2008} dwarfs respectively. Our young stars have \ewhe\ above the inactive dwarfs, confirming that active stars feature enhanced \helium\ triplet absorption. The dashed line and gray shaded region show a linear fit and 3$\sigma$-range for the EW-$T_{\rm eff}$ relation. The \ewhe\ outliers (V1298 Tau and HD 283869) are excluded from the helium-$T_{\rm eff}$ fit. The chromospheric \helium\ triplet has no $T_{\rm eff}$ dependence, unlike the photospheric \calcium\ line, highlighting the difference in line formation. The different combinations of marker shape and color for our sample correspond to particular stars as designated in Table~\ref{tab:objects}. This plotting scheme is also used in Figures~\ref{fig:he_val_prot}, \ref{fig:he_ew_scatter}, \ref{fig:he_ew_excess_scatter_prot}, \ref{fig:intra_visit_scatter}.}
    \label{fig:he_val_teff}
\end{figure}

The top panel of Figure~\ref{fig:he_val_teff} shows the median \ewhe\ as a function of $T_{\rm eff}$ for each star in our sample, along with literature values of inactive and active dwarfs. Relative to stars of the same effective temperature, the young stars in our sample show \ewhe\ that is enhanced over inactive dwarf stars \citep{zarro1986} and comparable to active dwarf stars \citep{sanzforcada2008}. The correlation between \ewhe\ and $T_{\rm eff}$ is consistent with a flat slope. This agrees with previous literature studies, in which no dependence on $T_{\rm eff}$ has been found for either inactive or active dwarf stars within this temperature range. From the lack of temperature dependence in our sample, we conclude that the stellar atmospheric conditions that lead to the formation of the \helium\ triplet are roughly the same for late-F to late-K dwarfs with $\tau \le 1$~Gyr. Table~\ref{tab:helium} shows the \ewhe\ median value and variability metrics for our sample described in the following subsections.

\begin{deluxetable*}{cccccc}
\tablecaption{Overview of Helium EW Measurements\label{tab:helium}}
\tablehead{
\colhead{Object} & \colhead{med[EW$_{\rm He}$]} & \colhead{MAD[EW$_{\rm He}$]\tablenotemark{a}} & \colhead{MAD$_{\rm intr}$[EW$_{\rm He}$]\tablenotemark{b}} & \colhead{$\sigma_{D, \rm rel, red}$\tablenotemark{c,$\bigstar$}} & \colhead{$\sigma_{D, \rm rel}$\tablenotemark{d,$\bigstar$}} \\
\colhead{} & \colhead{m\AA} & \colhead{m\AA} & \colhead{m\AA}  & \colhead{\%} & \colhead{\%} }
\startdata
V1298 Tau &   $447.6 \pm 5.8$ &  $62.5_{-7.7}^{+8.2}$  & $61.5_{-7.8}^{+8.4}$   & $-$                    & $3.93_{-0.5}^{+0.54}$\\
K2-284    &   $315.8 \pm 4.7$ &  $27.9_{-6.4}^{+11.2}$ & $23.5_{-10.0}^{+12.0}$ & $2.76_{-1.47}^{+1.02}$ & $3.11_{-1.32}^{+1.58}$\\
TOI 2048  &   $316.2 \pm 2.1$ &  $13.5_{-3.1}^{+3.9}$  & $4.3_{-4.3}^{+8.2}$    & $<1.53$                & $0.52_{-0.52}^{+1.02}$\\
HD 63433  &   $315.5 \pm 1.4$ &  $ 8.0_{-1.9}^{+1.2}$  & $7.1_{-2.2}^{+1.4}$    & $0.79_{-0.31}^{+0.21}$ & $0.80_{-0.25}^{+0.15}$\\
HD 283869 &   $191.3 \pm 2.0$ &  $12.5_{-2.7}^{+4.1}$  & $8.0_{-8.0}^{+5.6}$    & $0.68_{-0.68}^{+0.80}$ & $0.93_{-0.93}^{+0.66}$\\
K2-136    &   $302.2 \pm 1.3$ &  $13.4_{-1.8}^{+2.1}$  & $8.6_{-4.0}^{+3.2}$    & $1.33_{-0.50}^{+0.40}$ & $1.14_{-0.52}^{+0.42}$ \\
K2-100    &   $349.6 \pm 2.4$ &  $21.0_{-3.7}^{+5.0}$  & $6.6_{-6.6}^{+10.8}$   & $<2.75$                & $0.62_{-0.62}^{+1.01}$\\
K2-101    &   $323.1 \pm 3.3$ &  $ 8.7_{-6.6}^{+9.2}$  & $<20.0$                & $1.55_{-1.55}^{+4.86}$ & $<2.54$ \\
K2-102    &   $303.3 \pm 4.2$ &  $13.9_{-3.1}^{+10.8}$ & $<18.7$                & $<3.85$                & $<2.42$ \\
K2-77     &   $309.0 \pm 2.6$ &  $14.2_{-3.1}^{+3.8}$  & $<13.1$                & $<1.69$                & $<1.60$ \\
\enddata
\tablenotetext{}{$\bigstar$ Important note for interpretation: The intrinsic relative absorption depth variability, $\sigma_{D,\rm rel}$, represents the 1$\sigma$ amplitude of the intrinsic stellar \helium\ triplet variability. Thus it should not be directly used as a detection limit for exospheres. We suggest adopting 1.7-times this value as an estimated one-sided 95th percentile detection limit.}
\tablenotetext{a}{\ewhe\ time series variability, calculated as described in Section~\ref{sec:variability}.}
\tablenotetext{b}{The estimated intrinsic \ewhe\ variability, calculated as described in Section~\ref{sec:variability}. K2-101, K2-102, and K2-77 measurements are 90th percentile upper limits.}
\tablenotetext{c}{The intrinsic relative absorption depth variability of the \helium\ triplet's red component, as described in Section~\ref{sec:ew_to_depth_var}. V1298 Tau does not have a value because the feature is rotationally unresolved.}
\tablenotetext{d}{The intrinsic relative absorption depth variability of the full \helium\ triplet, as described in Section~\ref{sec:ew_to_depth_var}. In Figure~\ref{fig:exosphere_comp} we compare these values with literature helium exosphere measurements listed in Table~\ref{tab:lit_he_exospheres}.}
\end{deluxetable*}

The EW values of the nearby \calcium\ line are shown in the bottom panel of Figure~\ref{fig:he_val_teff}. There is a strong decrease in strength with increasing $T_{\rm eff}$, and a turnover at cooler temperatures. This highlights the difference in the line formation pathways between the chromospheric \helium\ triplet and the largely photospheric \calcium\ line. The \calcium\ EW-$T_{\rm eff}$ trend we find follows predictions from the photospheric PHOENIX spectral models \citep{husser2013}.

\begin{figure}
    \centering
    \includegraphics[width=\columnwidth]{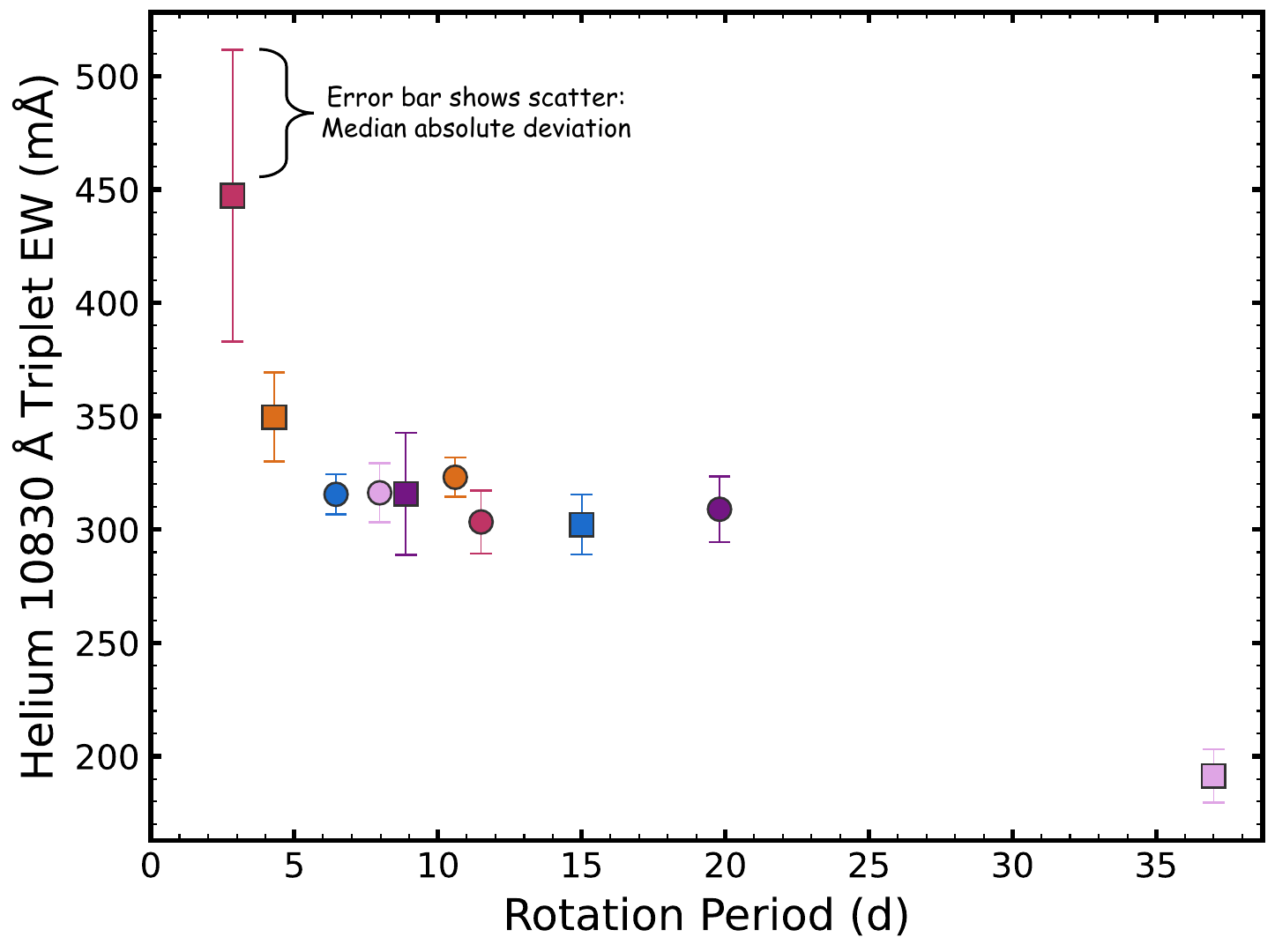}
    \caption{\ewhe\ as a function of stellar rotation period for our sample. The error bars are the median absolute deviation in \ewhe\ to show the amplitude of variability as compared to the \ewhe\ value (note this is not the uncertainty on the median \ewhe). \helium\ triplet absorption shows a clear morphology with the stellar rotation period: more absorption at shorter rotation periods, leading to a plateau below $P_{\rm rot} \sim 20$~d, before decreasing at the longest rotation period. This is indicative of the \helium\ feature's nature as an activity sensitive line: stars with shorter rotation periods have more intense high-energy coronal radiation and denser chromospheres, both of which help to form the stellar \helium\ feature.}
    \label{fig:he_val_prot}
\end{figure}

However, there are two outliers in the \ewhe-$T_{\rm eff}$ plane amongst our sample. This is likely explained by the stellar activity level, which can affect the population of metastable helium and therefore the absorption strength. One of the outliers is V1298 Tau, which is by far the youngest star in our sample ($\tau \sim 25$~Myr). The other is HD 283869, which is a peculiarly inactive Hyades member \citep{vanderburg2018}. As a proxy for activity level, we plot \ewhe\ as a function of stellar rotation period in Figure~\ref{fig:he_val_prot}. There is a clear variation in the absorption strength with respect to the rotation period: it is very strong at the fastest rotation, decreases through $\sim5$~days, plateaus between $5$ and $\sim20$~days, and then has a significant decrease in strength by $\sim35$~days. We note for comparison that the photospheric \calcium\ line does not show a rotation period dependence after removing the line's $T_{\rm eff}$ dependence. 

It is unsurprising that V1298 Tau would have the strongest \helium\ absorption strength in our sample because it is the most active due to its youth and rapid rotation. V1298 Tau is known to flare \citep{david2019a,david2019b,vissapragada2021,feinstein2022}, has a photometric variability amplitude indicative of a large spot filling fraction \citep{david2019a}, and potential spectroscopic indications of spots surrounded by facular regions \citep{feinstein2021}. All of these can enhance absorption in the \helium\ triplet, as well as lead to greater variability in its absorption strength (see Section~\ref{sec:variability}). 

K2-100 has the second highest \ewhe\ in our sample, which is unsurprising as it is also the second fastest rotator. This enhanced absorption strength is despite its older age, and may indicate that the structure of rapidly rotating stars introduces departures from models of chromospheres based on the Sun \citep[e.g.][]{andretta1995}.

HD 283869 has a long rotation period (nearly double that of the next slowest rotator in our sample), and weaker activity than other Hyades members \citep[from Ca~\textsc{ii}~HK and H-$\alpha$;][]{vanderburg2018}. Given the strong dependence of the \helium\ absorption strength on activity, it follows that our data would show weak absorption relative to more rapidly rotating stars. HD 283869 still resides at the upper envelope of the inactive dwarf \helium\ absorption sequence, so it may represent a transition between the active and inactive populations. Unfortunately, we do not observe stars with rotation periods between 20 and 35 days, which would help to clarify the nature of the weakening absorption (e.g. a gradual decline or a sharp discontinuity). Despite its lower \ewhe\ and slower rotation, HD 283869 is the same mass as fellow Hyades member K2-136. Surface gravity cannot explain the difference in absorption strength for this case.

While our sample does not cover a large range in $T_{\rm eff}$ or age, the stellar high-energy spectral distributions should still be changing across our targets. Differences in high-energy radiative output (such as in X-rays) would change the metastable helium population level if the PR mechanism dominates, in turn changing the \helium\ triplet absorption strength. Not all of our sample stars have literature X-ray observations, but given their rotation periods and masses they almost all likely rotate slower than the saturated X-ray luminosity regime \citep{pizzolato2003,douglas2014,nunez2015}. This means that the stellar X-ray luminosities should be decreasing with increasing rotation period in our sample. Our results showing a lack of dependence on $T_{\rm eff}$ or rotation period (within 5 to 20 days) reinforces the conclusion from \citet{sanzforcada2008} that CE is more important than the PR mechanism in driving the population of metastable helium for active stars. This is likely due to active stars having hotter and denser chromospheres. In the \ewhe-$P_{\rm rot}$ plateau, CE dominates any decrease in PR excitation from decreasing high-energy coronal radiation, leading to no change in the EW. 

Interestingly, V1298 Tau has a median \ewhe\ larger than the maximum predicted by models with active region filling factors of unity from \citet{andretta1995}, even including some epochs with values greater than 50\% higher. In these models, the resulting predicted \ewhe\ is set by the density and temperature structure of the chromosphere. \ewhe\ increases as activity increases, which is parameterized with the chromospheric density. Once the density is high enough, collisional de-excitation controls the metastable helium population and drives the line into emission, leading to a turnover in \ewhe\ \citep{andretta2017}.

It is possible that V1298 Tau’s youth, even relative to the other stars in our sample, alters chromospheric conditions to further heighten the \helium\ triplet absorption strength. For example, there are activity-driven phenomena not considered in these models, such as a constant level of flaring or eruption, that could be unique to particularly young stars. These conditions would alter the metastable helium population, and thus the models of \ewhe. It is also important to point out that V1298 Tau has a lower $T_{\rm eff}$ than the models, which \citet{andretta1995} explicitly state may be an issue for K-type stars.

Also, V1298 Tau has a lower surface gravity due to its youth that may lead to a larger chromospheric scale height with reduced density. This may seem counter to our earlier argument that young stars have higher \ewhe\ because of {\it denser} chromospheres. However, we posit that V1298 Tau has an extended, more rarefied upper chromosphere near the transition region above denser chromospheric layers. This can be illustrated with the VAL C model in Figure 4 of \citet{andretta2017}, where a decrease in the upper region density corresponds to an increase in the predicted maximum \ewhe. In this case, the column density of metastable helium increases to produce a larger \ewhe\ before reaching the volume density that initiates collisional control of the population. With this lower density, coronal EUV radiation would also be able to penetrate deeper into the chromosphere to raise the contribution of the PR mechanism to population metastable helium. A combination of density and radiation conditions in the chromosphere would thus lead to elevated \ewhe\ in stars of V1298 Tau’s age.

Regardless of the speculated cause, our results imply that the chromospheres of young and active stars diverge from models used for main sequence solar-type stars. In the case of the youngest and most active stars, like V1298 Tau, conditions likely significantly diverge from previous literature models. Between an altered, denser chromospheric structure, rapid rotation, and more intense high energy radiative environment, the \helium\ triplet’s absorption is much greater for active stars than inactive stars.


\subsection{Amplitude of variability in the \helium\ triplet's absorption strength}\label{sec:variability}

\begin{figure*}
    \centering
    \includegraphics[width=\columnwidth]{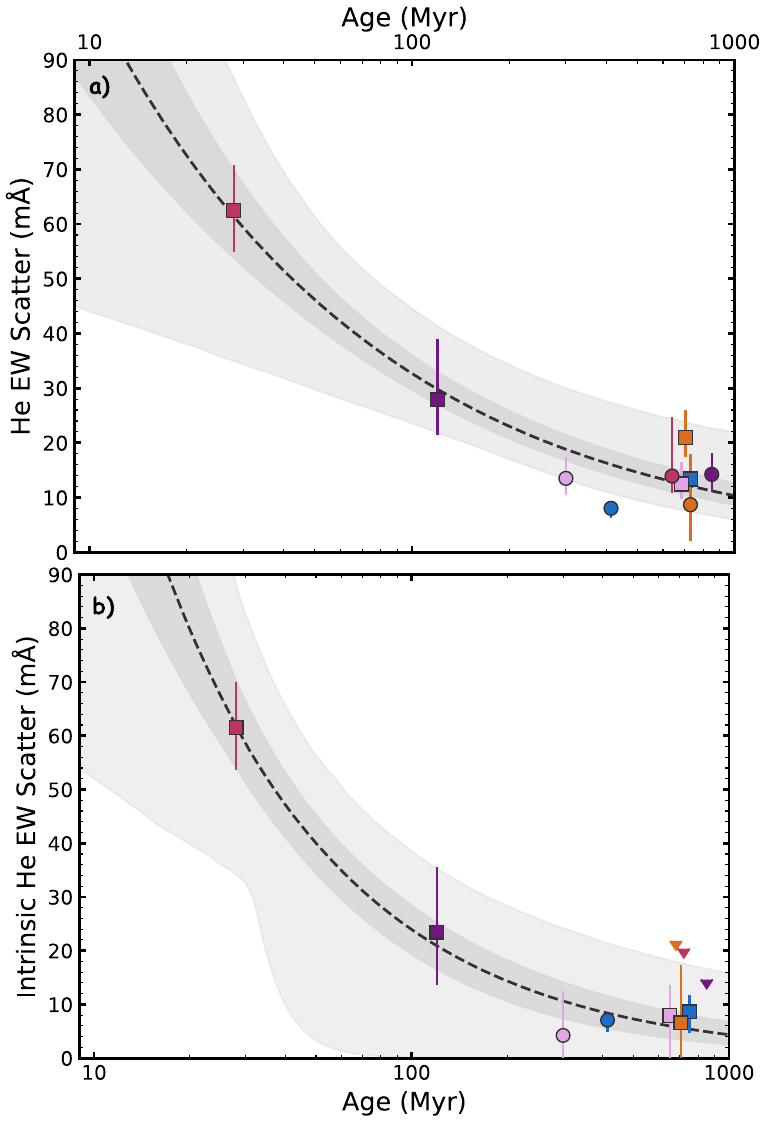}
    \caption{Long-term time-averaged \ewhe\ variability as a function of age. The dashed lines show power law fits to the data, with the shaded regions denoting the 1-$\sigma$ (darker gray) and 3-$\sigma$ (lighter gray) intervals. The plotted ages for the Hyades and Praesepe members are artificially spread to avoid overlapping points. \textit{Top panel:} The variability shown is the median absolute deviation of the \ewhe\ time series, scaled by 1.4826 to be statistically equivalent to the Gaussian standard deviation. Using the MAD accounts for outliers from low spectral S/N or poor spectral fitting. The error bars are generated from bootstrap sampling of the data (see text for further explanation). The variability decreases with age, plateauing above 300 Myr at $10-15$~m\AA. The variability increases significantly at younger ages, but the relation is ill-defined due to only having two targets with $\tau \le 120$~Myr. \textit{Bottom panel:} The plotted MAD is an estimate of the intrinsic variability in \ewhe, quantifying variability that cannot be accounted for with measurement error (see text for further explanation). Only upper limits are measured for 3 targets, for which the 90\% upper limits are denoted by the triangle marker. The variability-age relation persists, showing that the relation is intrinsic to the stars. The objects at the oldest ages are in agreement with intrinsic variability between $5-10$~m\AA, with 6 of the 8 older targets in agreement with no intrinsic variability due to the magnitude of their \ewhe\ measurement uncertainties.}
    \label{fig:he_ew_scatter}
\end{figure*}

\begin{figure}
    \centering
    \includegraphics[width=\columnwidth]{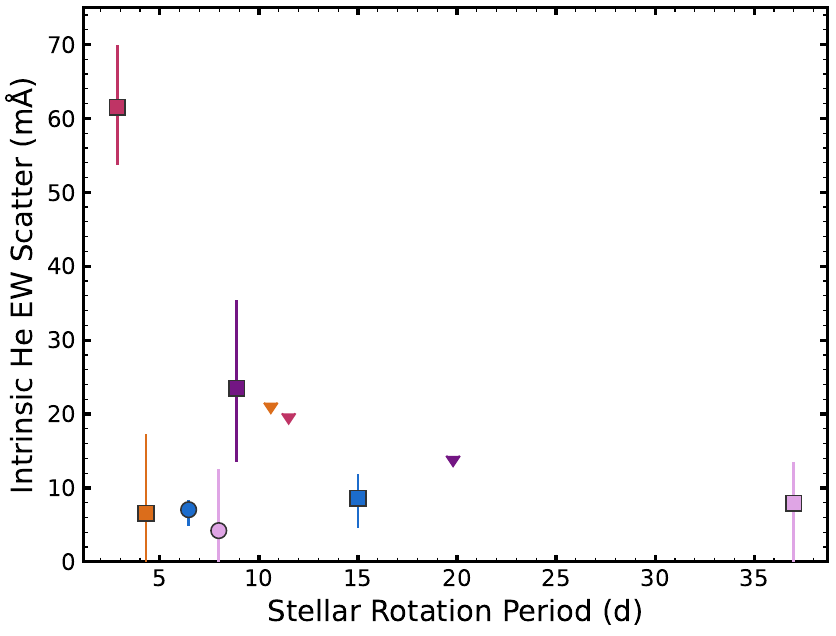}
    \caption{Intrinsic variability of \ewhe\ as a function of stellar rotation period, following the calculation described in the text and also shown in Figure~\ref{fig:he_ew_scatter}. There is no relation between the variability and rotation period, which we use as a proxy for activity level. This is unlike for the EW itself, which correlates with the rotation period. The two objects with high variability (V1298 Tau and K2-284) are younger than the rest of the sample.}
    \label{fig:he_ew_excess_scatter_prot}
\end{figure}

Figure~\ref{fig:he_ew_scatter} shows that the stars in our sample span a wide range of \ewhe\ variability, with very high variability in youth that decreases expeditiously towards older ages. We quantify the variability using the median absolute deviation (MAD), scaled by 1.4826 to be statistically equivalent to the Gaussian standard deviation. There is variability in the \ewhe\ value both with (top panel) and without (bottom panel) accounting for measurement uncertainty, which will inflate the measured MAD, indicating that at least some changes in the \helium\ triplet absorption strength are intrinsic to the stars.

In the top panel of Figure~\ref{fig:he_ew_scatter} we show the MAD of the \ewhe\ time series for each star. To measure the uncertainty of our measurement of the MAD, we perform bootstrap sampling of each target’s \ewhe\ time series. For a target with $N$ measurements, we generate a simulated time series by sampling $N$ \ewhe\ values with replacement and calculate its MAD. We repeat this process 500000 times, and adopt the median and 16th-84th percentile ranges as the value and uncertainty in the MAD, respectively. The relationship between \ewhe\ and age is well described by a power law, which we fit to our data and show in the figure. The variability is highest at the youngest ages ($\sim25$~Myr), roughly halves by $\sim100$~Myr, before plateauing at ages older than $\tau \sim 300$~Myr.

However, some of this measured variability is introduced by the uncertainty of our \ewhe\ measurements. To estimate the true {\it intrinsic} variability of the \helium\ triplet absorption strength, we compute a MAD quantity with the measurement uncertainty deconvolved and show it in the bottom panel of Figure~\ref{fig:he_ew_scatter}. For each target, we first generate a simulated time series with no intrinsic variability by sampling for each data point from a Gaussian distribution with a mean of 0 and a standard deviation of the individual measurement uncertainty. The MAD of this time series, which we call MAD$_{\rm meas}$, encapsulates just the measurement uncertainty. We then sample one of the 500000 bootstrapped time series MAD values described in the previous paragraph to represent the combined intrinsic and measurement variability, called MAD$_{\rm full}$. The intrinsic variability is then estimated as:

\begin{equation}\label{eq:intrinsic_mad}
	{\rm MAD}_{\rm intr} = \sqrt{{\rm MAD}_{\rm full}^2 - {\rm MAD}_{\rm meas}^2}
\end{equation}

\noindent We repeat this 500000 times to build a distribution of intrinsic variability estimates. If the measurement-only variability is larger than the combined variability, we set the MAD$_{\rm intr}$ to 0, as such a time series agrees with no intrinsic variability. We adopt the median and 16th-84th percentiles of the MAD$_{\rm intr}$ distribution as the value and uncertainty. If the median value for a target is 0, we adopt the 90th percentile as an upper limit on the intrinsic variability. 

A power law also describes the relationship between intrinsic \ewhe\ variability and age, with a plateau below 10~m\AA\ at ages above 300 Myr and increasing variability below 120 Myr. Of the 8 stars in the older age range, 6 of them have intrinsic variability estimates that agree with 0, so the plateau above 300 Myr may agree with no intrinsic variability. The 3-$\sigma$ range of the fit encompasses the scenario where there is no intrinsic variability at older ages with a sharp, immediate increase at the youngest ages. This is not the preferred fit, though, and 2 of the older stars in our sample do have resolved intrinsic variability (K2-136 and HD 63433). 

We make particular note of HD 63433, which is by far the brightest star in our sample, leading to the highest typical spectral S/N and therefore the lowest typical measurement error on \ewhe. HD 63433 may therefore be the exemplar of intrinsic \ewhe\ variability in this age range, disputing the lack of intrinsic variability from our upper limit measurements but confirming that the level of variability is relatively low. The 1-$\sigma$ range of intrinsic \ewhe\ variability measurement for HD 63433 is 4.9 to 8.4~m\AA, which corresponds to 1.5 to 2.7\% when normalized by its median \ewhe\ value. For our highest quality spectra we can measure the intrinsic variability to a precision of roughly 0.5\%. 

We tentatively suggest that this roughly 2\% intrinsic long-term average \ewhe\ variability is likely representative of the typical variability in the 300 Myr to 1 Gyr age range, although the multiple measurements that agree with 0 show that some targets may have no intrinsic variability and that more monitoring at higher S/N is necessary to resolve the intrinsic variability.

We do want to address the stars that agree with no intrinsic variability, and why it may or may not be indicative of an actual lack of variability. Of the 8 stars in this older age range, 6 of them have intrinsic variability estimates that agree with 0. We only determine upper limits for 3 of the stars, meaning that any intrinsic variability is below the contribution of measurement error. Two of the three stars with upper limits, K2-101 and K2-102, are relatively faint with the fewest number of observations in our sample, which would lead to larger measurement uncertainty. The other star with an upper limit, K2-77, is also on the faint end of our sample in addition to being the oldest, so it may be indicative of a true lack of intrinsic variability. For 3 other stars (K2-100, HD 283869, and TOI 2048), we resolve an intrinsic variability measurement that agrees with 0 within 1-$\sigma$. K2-100 is not surprising, as its broad lines from rapid rotation lead to larger \ewhe\ measurement uncertainty than is typical. HD 283869 is by far the least active star in our sample, so it agreeing with no intrinsic variability is expected. TOI 2048 is slightly more surprising, although it has the shortest survey baseline with half of its observations taken within a single month which may lead to less variability being captured. 

The variability we find at $\tau \gtrsim 300$~Myr is comparable to the field M-dwarfs studied by \citet{fuhrmeister2020}. The median variability across their sample is $\sim12$~m\AA\ (also quantified with MAD), which is similar to the plateau in variability at older ages in our sample (top panel of Figure~\ref{fig:he_ew_scatter}). While the strength of \helium\ absorption decreases through the M-dwarf regime, this result shows that above ages of $300$~Myr, the activity-induced variability does not have a significant spectral type dependence through SpT $\sim$ M3. The young V1298 Tau and K2-284 do have larger variability amplitudes than the early M-dwarfs. K2-284 has variability comparable to the late M-dwarfs and V1298 Tau has variability larger than all but a few of the M-dwarfs in the CARMENES sample. Activity-induced variability is significant at young ages and low effective temperatures, being the most significant at the youngest ages regardless of SpT.

While variability increases significantly at younger ages, there is no correlation between variability and rotation period. This stands in contrast to the dependence of \ewhe\ on rotation period (see Figure~\ref{fig:he_val_prot}). Figure~\ref{fig:he_ew_excess_scatter_prot} shows the intrinsic variability of \ewhe\ as a function of stellar rotation period, where stars with ages above 300 Myr either have comparable scatter to the most slowly rotating stars or only upper limits for the intrinsic variability. Rotation is a direct proxy for activity level, so we conclude that there must be a non-rotation activity effect that drives variability in the helium absorption strength. The two outliers in Figure~\ref{fig:he_ew_excess_scatter_prot} are both young, so age may be a factor. V1298 Tau is the most significant outlier, while being the youngest and most rapidly rotating star in our sample. Its helium variability may therefore be driven primarily by its rotation. While the other outlier, K2-284, is also young, it has a typical rotation period for our sample but a later spectral type \citep[late-K to M0, rather than mid-K or earlier;][]{david2018,dressing2019}. It may be more active than earlier-type stars at the same rotation period, resulting in greater helium variability.

This lack of clear relation between helium variability and rotation implies that there is a combination of effects on the stellar activity level that drive changes in the helium absorption strength, such as from age, rotation, and stellar structure (spectral type). We conclude that the \ewhe\ variability is set by activity evolution. Examples of this would be from acute changes in the \helium\ absorption strength \citep[e.g. from flares;][]{vissapragada2021}, short-term variability from rotational modulation of active regions \citep[e.g. spots and plage, as on the sun; e.g.][]{brajsa1996}, and long term activity cycles \citep[as seen for solar-type stars in other chromospheric lines such as Ca \textsc{ii} H and K; e.g.][]{baliunas1995,borosaikia2022}. Further work studying the relationship between activity and rotation period at older ages, such as by studying fast rotating tidally locked binaries, would help disentangle the complicated connection between activity level and the stellar dynamo.

We show that the long-term time-averaged variability in \ewhe\ is stable and relatively low for stars with $\tau \gtrsim 300$~Myr, and cannot rule out the absence of intrinsic variability for most of our targets in that age range. This means that even observations separated by months to years might have a comparable \ewhe\ baseline, although there are physical mechanisms that may drive variability on shorter timescales; the relative timing of observations and variability will be a crucial constraint on exosphere detectability. Stars older than 300 Myr seem to behave like field stars, but we find high-amplitude statistically significant intrinsic variability for stars with $\tau \lesssim 120$~Myr. Observations of stars with ages between $50-300$~Myr are needed to fully map the \ewhe\ variability-age dependence. Such measurements will become available as NIR RV surveys of young stars continue \citep[e.g.][]{tran2021}.

\subsection{Timescale of variability in the \helium\ triplet}\label{sec:timescales}

\begin{figure}
    \centering
    \includegraphics[width=\columnwidth]{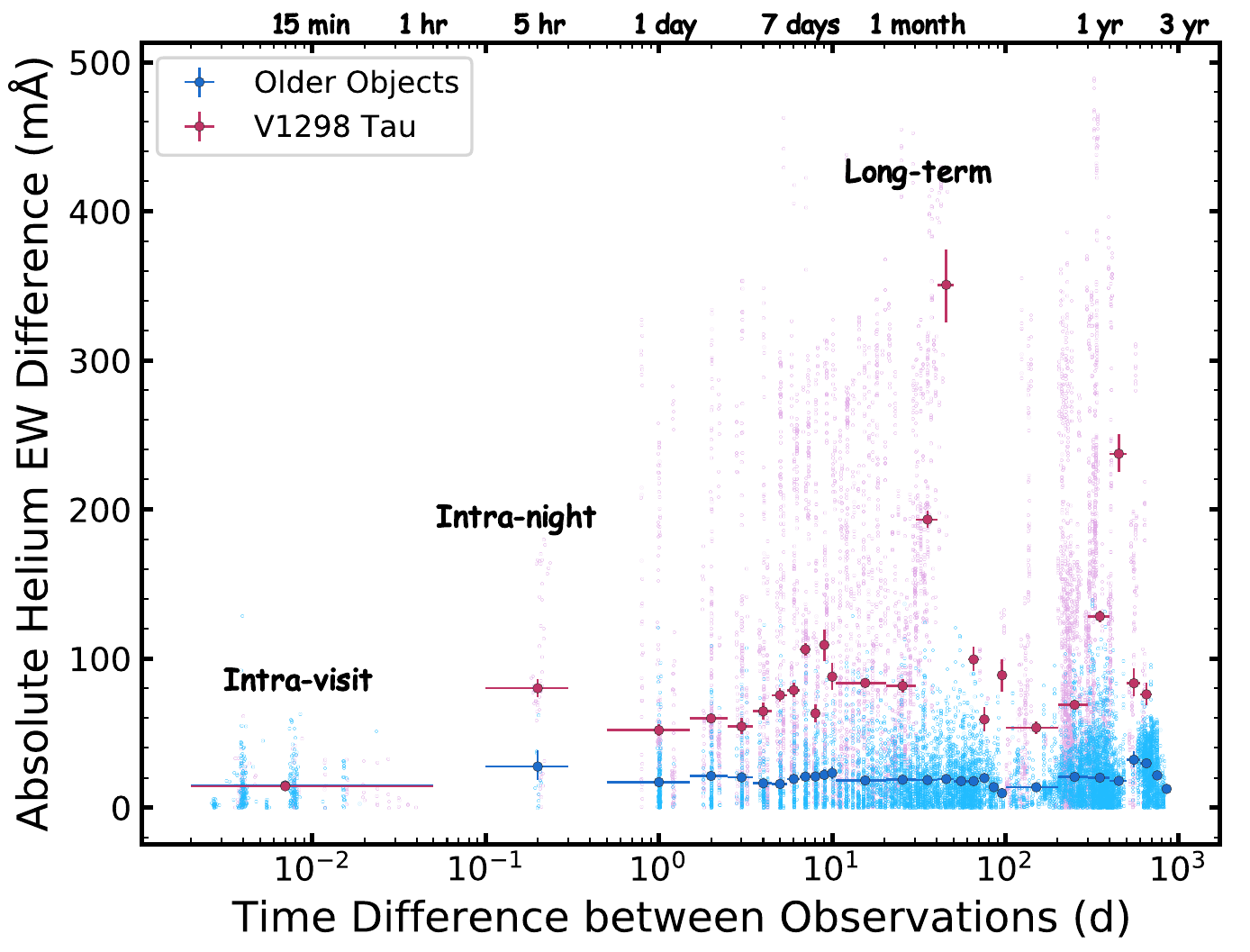}
    \caption{The \ewdiffspectrum\ of the \helium\ triplet time series for our targets. We compute the pairwise difference in the \ewhe\ for each observation of a particular target as a function of the time separating the observations. We stack the \ewdiffspectra\ for all targets besides V1298 Tau, as they are older and have lower long-term average variability. The small background points show all pairwise differences. We generate 10000 \ewdiffspectra\ accounting for measurement uncertainty on individual \ewhe, and the large foreground points show the median of these \ewdiffspectra\ samples baseline-binned to reflect the typical variability at each timescale. The x-axis error bars show the timescale bin extents, and the y-axis error bars show the 16th-84th percentile error on the median \ewdiffspectrum. The variability is smallest at the shortest timescales, increases after only a few hours, and plateaus out to separations of years.}
    \label{fig:ew_diff_spectrum}
\end{figure}

Our above analysis of the \helium\ triplet absorption variability represents the time-averaged variability over the full observation baseline of our targets, spanning minutes through years. While this provides insight into the typical variability amplitude of the \helium\ triplet, it does not reveal the timescale over which this variability occurs. As exosphere observations are inherently temporal, the timescale of variability is crucial to assess the feasibility of young exosphere detection. Most exosphere detection programs include \oot\ observations on the same night as the transit, which would be affected by stellar variability on timescales of hours. Our long-term variability analysis shows the limits of comparing observations over full observing seasons (weeks to months to years). We must investigate the timescale of variability in \ewhe\ to assess \oot\ observation comparison at shorter time baselines.

To quantify the \helium\ triplet variability as a function of the time baseline between observations we compute an \ewdiffspectrum. For each target, we calculate the difference in the \ewhe\ for each pair of observations as a function of the time separation of the observations, and we show the result in Figure~\ref{fig:ew_diff_spectrum}. The \ewdiffspectra\ for all targets besides V1298 Tau are stacked to increase precision, because they have roughly the same amount of long-term average variability and because these ``older" stars appear to act similarly. V1298 Tau is plotted separately to highlight the increased variability at the youngest ages. To account for measurement uncertainty, we generate 10000 mock time series for each object using the \ewhe\ values and errors and compute the 10000 corresponding \ewdiffspectra. The large points plotted in Figure~\ref{fig:ew_diff_spectrum} are the median of the 10000 sample \ewdiffspectra, and the error bars are the 16th and 84th percentile values to represent the error on the median difference. The x-axis error bars show the timescale bin extents.

At the shortest timescales ($\Delta t\la1$~hr), the variability is lower than for longer time baselines, and the same for both the ``older" objects and V1298 Tau. While variability may still be a significant issue depending on the exosphere parameters, it is smallest on the shortest timescales regardless of age and may be most affected by acute changes in \helium\ triplet absorption (e.g. flaring). For the ``older" objects, the variability increases slightly beyond $\Delta t\sim5$~hr, and is then relatively stable at longer timescales. The variability for V1298 Tau follows a similar morphology, but has a \textit{much} larger amplitude, even on the same night, which is expected given its larger long-term variability and the increased volatility in activity at young ages.

The typical EW variability increases significantly even after just a few hours, represented by the second binned data point at 5 hours in Figure~\ref{fig:ew_diff_spectrum}. The only two targets with observations covering this baseline are V1298 Tau and K2-100, both of which have relatively short rotation periods. It is important to note that there were only two nights for which K2-100 was observed on both tracks available for the HET (as opposed to 7 nights for V1298 Tau), so the second binned data point for the older objects must be interpreted with caution. It is possible that the heightened variability on this timescale is from active regions of varying \helium\ absorption strength beginning to rotate in and out of view. This may also be capturing acute changes in the \helium\ absorption strength due to flaring (particularly for V1298 Tau). This timescale of increased variability might be due to the lifetime of the helium metastable state, which is 2.2 hours \citep{drake1971}. The changing \helium\ absorption strength on this relatively short timescale of $\Delta t\sim5$~hr may represent de- and re-population of the helium metastable state between observations. Above timescales of $\Delta t\sim5$~hr, the EW difference plateau represents variability in the \helium\ absorption strength (e.g. from active regions, flares, stochastic activity) that is averaged out over multiple epochs.

\subsubsection{Short timescales: intra-night variability}\label{sec:intranight_var}

\begin{figure}
    \centering
    \includegraphics[width=\columnwidth]{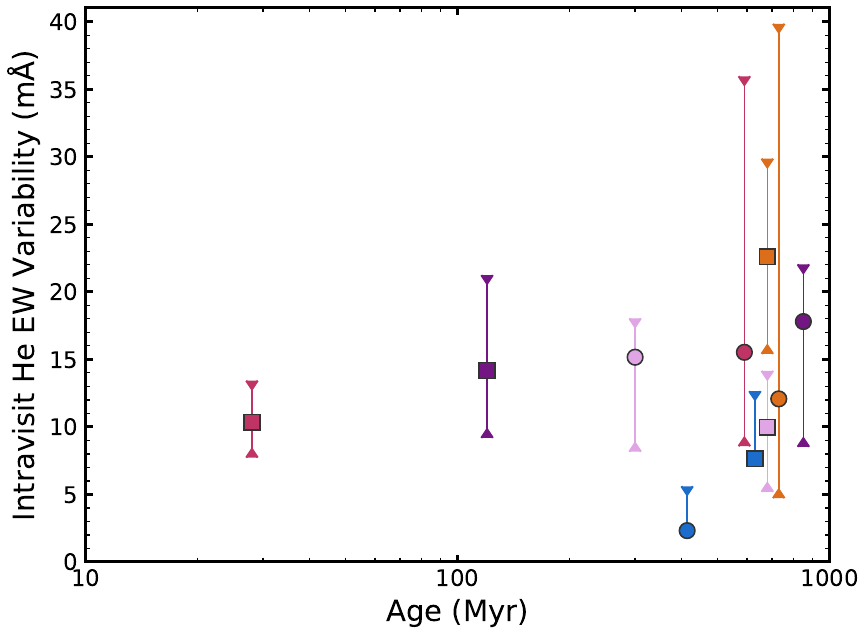}
    \caption{Intra-visit variability of \ewhe\ as a function of age, to demonstrate variability within a 30 minute baseline. The plotted ages for the Hyades and Praesepe members are artificially spread for clarity. The markers show the median \ewhe\ difference for pairs of observations within the same visit for each object. The lines bracketed by triangles are not error bars. They show the 90th percentile range of intra-visit variability measurements that would be found if we assume no intrinsic variability, only measurement error. The measured intra-visit variability is in agreement with no intrinsic variability for 8 of the 10 objects, and just in slight disagreement for the other 2. At these shortest timescales, the variability measurement is independent of age and can be explained solely by measurement error without any intrinsic \ewhe\ variability. This gives confidence that detecting even the youngest exospheres is possible with comparison observations directly surrounding transit.}
    \label{fig:intra_visit_scatter}
\end{figure}

We can also assess variability on the shortest timescales by inspecting measurements taken on the same night. An observation of a target, which we call a visit, is composed of multiple consecutive exposures (normally 3). This provides multiple spectra separated by $5-10$ minutes. Figure~\ref{fig:intra_visit_scatter} shows the intra-visit \ewhe\ variability, representing timescales of $\la30$~minutes, as a function of age. We quantify the intra-visit variability by computing the absolute EW difference between each pair of observations in a single visit, and then present the median pairwise difference for each target. This is directly comparable to exosphere detections which are presented as measurements of excess absorption depth, which are essentially differences in EW.

Compared to the long-term time-averaged variability (Figure~\ref{fig:he_ew_scatter}), the intra-visit variability has no dependence on age, and has similar or smaller amplitude to the MAD of the \ewhe\ time series. These conclusions are in agreement with those drawn from the \ewdiffspectra\ about variability on the shortest timescales (Figure~\ref{fig:ew_diff_spectrum}, Section~\ref{sec:timescales}).

The intra-visit variability measurements represent both measurement error and the intrinsic variability of \helium\ absorption strength. To determine the significance of the measured variability, we estimate the intra-visit variability produced solely by the measurement uncertainty. For each target, we generate 5000 \ewhe\ time series wherein the \ewhe\ value is assumed to be constant within a visit and the individual measurements are sampled using the measurement uncertainty. We then calculate the intra-visit variability for each simulated time series following the procedure described previously. The vertical lines bracketed by triangle markers plotted in Figure~\ref{fig:intra_visit_scatter} show the smallest range of intra-visit variability values that include 90 percent of the simulations, denoting the range of values consistent with no intrinsic variability.

Of the 10 objects in our sample, 8 have measured values for their intra-visit variability that agree with no intrinsic intra-visit variability. The other 2 objects (HD 63433 and K2-136) just barely do not agree. This result agrees with all objects having no intrinsic intra-visit variability. This is promising for the detection of young exospheres if \oot\ observations are taken directly surrounding transit. The intra-visit variability measurements of the stars with $\tau \ga 300$~Myr are comparable in amplitude to their long-term time-averaged variabilities, when not deconvolving the measurement uncertainty. However, while some objects in this age range have measurable statistically significant intrinsic long-term time-averaged variability, {\it all} objects have intra-visit variability in agreement with only being caused by measurement quality. This reinforces the conclusion that variability on such a short timescale should be negligible, and the notion that higher S/N from stacked spectra taken across the transit will increase the reliability of exosphere observations.

However, there still could be increased variability on short timescales. The \helium\ triplet strength increases from flares \citep{fuhrmeister2020}, which act on timescales of minutes to hours. \citet{vissapragada2021} even found a linear increase in the \helium\ triplet strength of V1298 Tau in the decay phase of a flare, potentially indicating the line's temporal response. The observation was during a transit of V1298 Tau c, though, so the increase could have been due to either the flare or an exosphere. Other acute active phenomena could change the \helium\ triplet strength, such as spot/plage rotation, chromospheric network variability, winds, and mass loss. While variability is small on timescales that cover \oot\ observations directly surrounding transit, even agreeing with no intrinsic variability on timescales within 30~minutes, care must still be taken when interpreting \helium\ triplet transit observations in the context of stellar activity.

\subsubsection{Intermediate timescales: intensive observing campaigns}\label{sec:intensive_campaigns}

We further investigate variability in the \helium\ absorption strength at intermediate timescales using the intensive campaigns that were taken for three targets in our sample: V1298 Tau, K2-136, and K2-100. These intensive campaigns include regular observations over a months-long span, with individual visits often only separated by a day. The high cadence of these data is useful to search for periodicity, unlike the sparse, years-long time series for most targets in our sample. More frequent observations are also more likely to catch acute changes in the \helium\ absorption such as from flares. We computed periodograms for each intensive time series to search for periodicity, and discuss these observations below.

\begin{figure}
    \centering
    \includegraphics[width=\columnwidth]{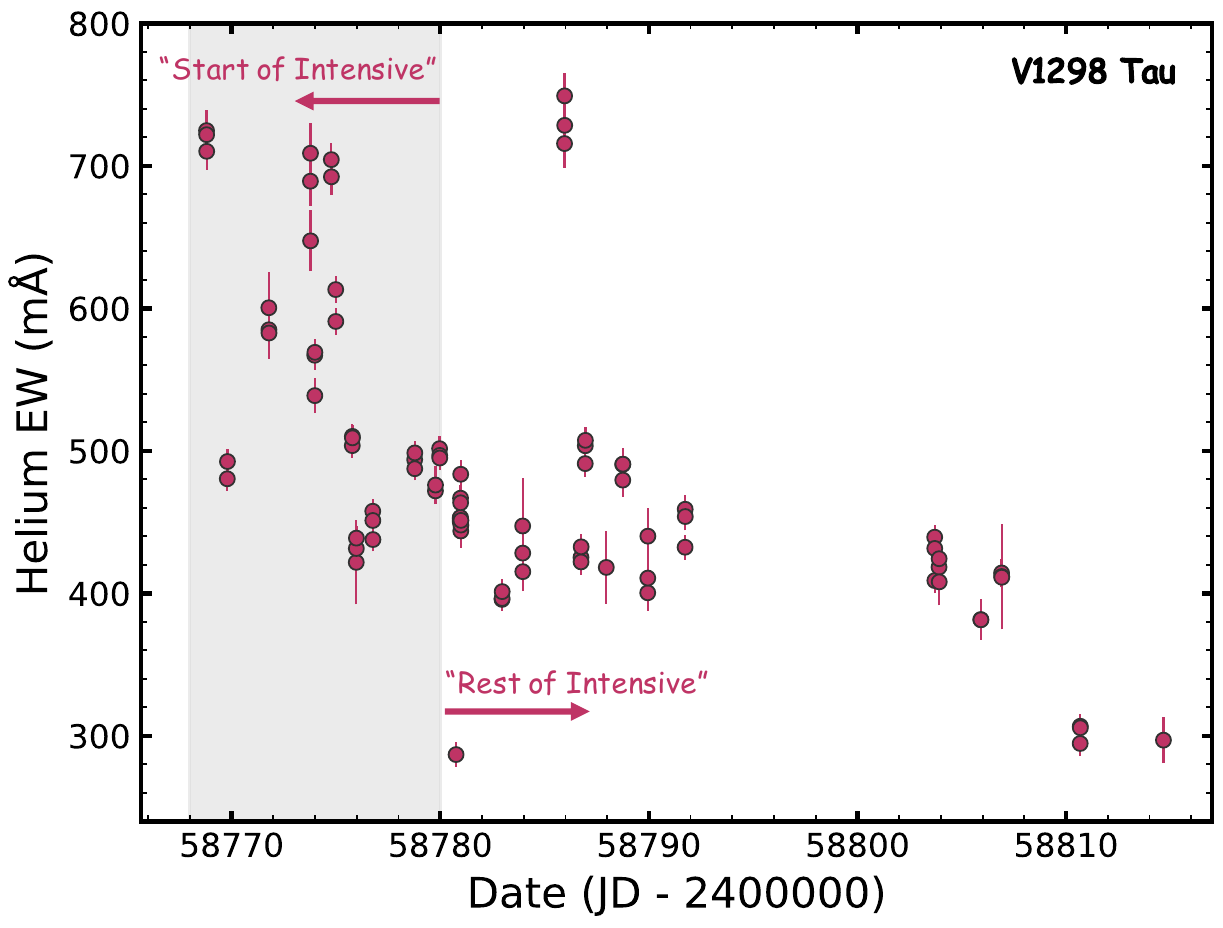}
    \caption{The \ewhe\ time series for the V1298 Tau intensive campaign. In the first week and a half of the campaign there is extreme variability in the EW value, which then steadies at a slightly lower value for the remaining time series coverage.}
    \label{fig:v1298tau_intensive}
\end{figure}

\textbf{V1298 Tau:} With the highest \ewhe\ variability in our sample, V1298 Tau provides an important opportunity to study significant absorption strength changes at high cadence. V1298 Tau is a unique system: it is a very young ($\tau \sim 25$~Myr) early K-dwarf, hosts 4 known transiting planets \citep{david2019a,david2019b}, is a promising candidate for follow-up atmospheric characterization of young planets, and is a useful probe of star and planet formation as a member of the older distributed stellar groups around Taurus \citep{krolikowski2021}. The intensive campaign for V1298 Tau comprises 30 visits (98 spectra) spanning 45 days, with a median visit separation of 1~day. There was a brief break in observations for 12 days in the middle of the campaign. This intensive campaign was taken in October-November of 2019 with the goal of measuring the mass of V1298 Tau b, which will be presented in a future paper. Figure~\ref{fig:v1298tau_intensive} shows the intensive \ewhe\ time series of V1298 Tau, which features significant structure. The \ewhe\ varies by almost $300$~m\AA\ in the first week and a half of the intensive campaign, before settling down to vary by $\sim100$~m\AA\ around a stable value, and then dropping further in the last few days. There is also an observation with greatly elevated \ewhe\ at JD $\sim2458786$ that may be indicative of a flare. 

\begin{figure}
    \centering
    \includegraphics[width=\columnwidth]{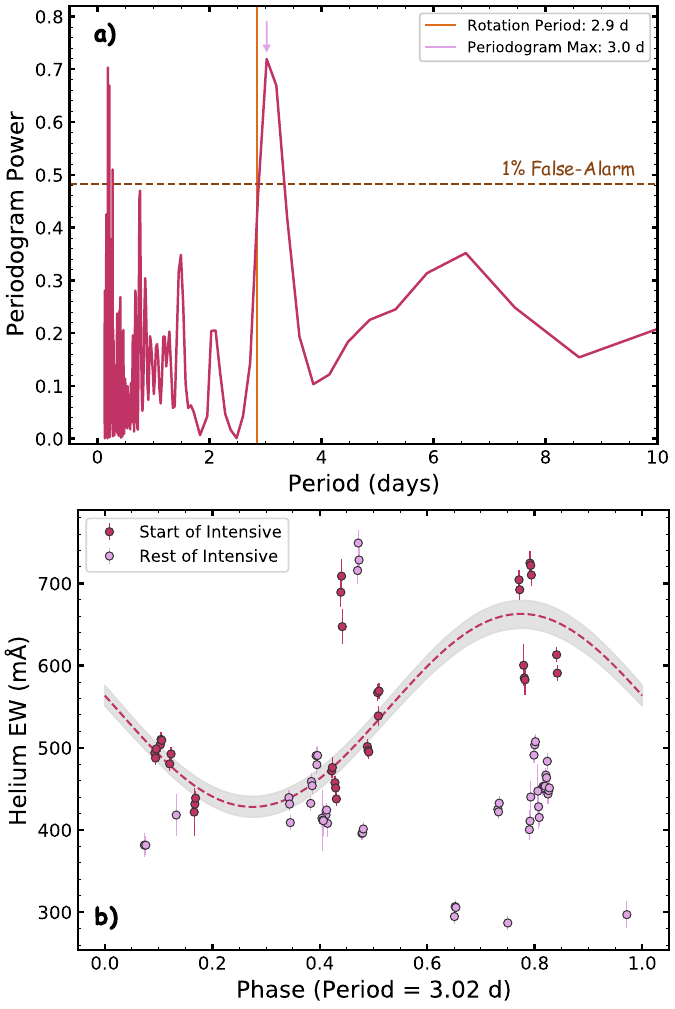}
    \caption{Periodicity analysis of the V1298 Tau intensive campaign. \textit{Top Panel:} The periodogram of the first week and a half of data from the intensive campaign, which features extreme variability. The peak of the periodogram ($P=3.02$~d) is very close to the rotation period of the star ($P=2.9$~d), and has power well above the 1\% false alarm probability level. \textit{Bottom panel:} The intensive campaign phased to the periodogram peak's period, shown for the first week and a half separately from the rest of the intensive time series data. The coherence at the start of the campaign is stark, while there is no apparent rotational modulation for the data from the remainder of the campaign.}
    \label{fig:v1298tau_periodicity}
\end{figure}

There is no significant periodicity across the entire time series, which may indicate that the variability in the \helium\ absorption strength comes from equal spatial distribution of active regions across the surface, such as extreme spot coverage and the chromospheric network. However, there \textit{is} significant periodicity near the stellar rotation period within the first week and a half of the campaign when the absorption varies wildly. The periodogram of the first week and a half of data is shown in the top panel of Figure~\ref{fig:v1298tau_periodicity}. The bottom panel of Figure~\ref{fig:v1298tau_periodicity} shows the intensive campaign phased to the period of peak periodogram power, and the data from the first week and a half is coherent unlike the rest of the time series.

This contradicts the conclusion from the entire intensive data set that variability is essentially smeared out across the stellar rotation. We posit that the large amplitude variation at the stellar rotation period may be from an extreme flaring or mass loss event that created a large but spatially concentrated bright spot on the stellar chromosphere, which would then show enhanced \helium\ absorption when it rotates into view. This highlights the volatility in \ewhe\ at such high activity levels, and warrants even further caution for planning time sensitive transit observations. 

\begin{figure}
    \centering
    \includegraphics[width=\columnwidth]{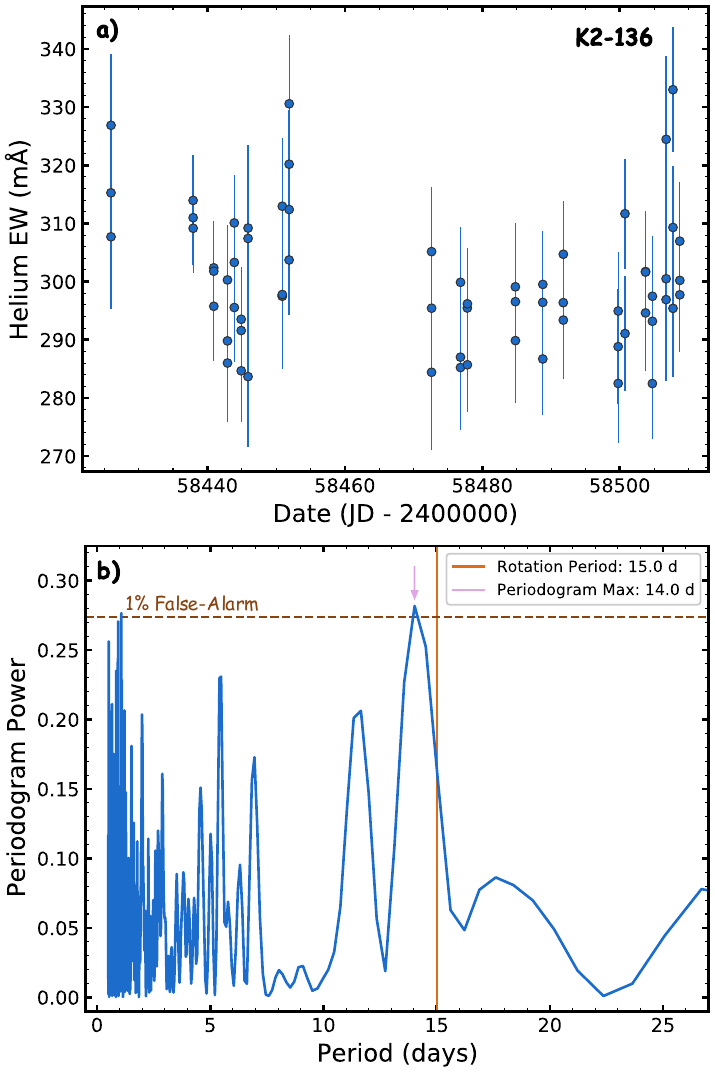}
    \caption{\textit{Top Panel:} The \ewhe\ time series for the K2-136 intensive campaign. The data is fairly flat, and most of the variability shown is within the individual EW measurement uncertainties. \textit{Bottom panel:} The periodogram for the full intensive campaign of K2-136, which shows a peak that is significant just above the 1\% false alarm probability level. The period of the peak ($P=14$~d) is close to the stellar rotation period ($P=15$~d). Given the power of the peak, the periodicity found here is taken to be inconclusive.}
    \label{fig:k2136_intensive}
\end{figure}

\textbf{K2-136:} K2-136 is a K5.5-dwarf Hyades member that hosts 3 transiting planets \citep{mann2018,ciardi2018,livingston2018}, and is representative of the older targets in our sample that have \helium\ variability in line with the field. It has a rotation period on the slower end of our sample's range (15 days), with 7 of the 10 stars having faster rotation. The intensive campaign for K2-136 comprises 22 visits (68 spectra) spanning 80 days, with a median visit separation of 2~days. There was a brief break in observations for 20 days in the middle of the campaign. This intensive campaign was observed soon after HPF's commissioning, and has been used as a standard reference by our team in analyzing HPF data. The top panel of Figure~\ref{fig:k2136_intensive} shows the intensive \ewhe\ time series of K2-136, which is fairly flat over the course of the observations. We generate mock time series assuming constant EW using each observation's measurement error to determine if the data variability is consistent with zero. The scatter for the data is $\sigma=11.3$~m\AA, which agrees with the simulated scatter assuming no intrinsic variability of $\sigma=10.3\pm1.0$~m\AA. It is unsurprising that an older, less active star would show less structure in the high cadence \ewhe\ measurements. The intensive campaign has periodicity near the stellar rotation period, shown by the periodogram in the bottom panel of Figure~\ref{fig:k2136_intensive}, but it has power just barely above the 1\% false alarm probability level. This potential rotational modulation is inconclusive, and we do not show the phased time series because it does not show clear modulation. The average EW during the intensive campaign is lower than for the rest of the sparse time series, which may be indicative of longer term (months to years-long) activity cycles manifesting in the \ewhe\ of K2-136.

\begin{figure}
    \centering
    \includegraphics[width=\columnwidth]{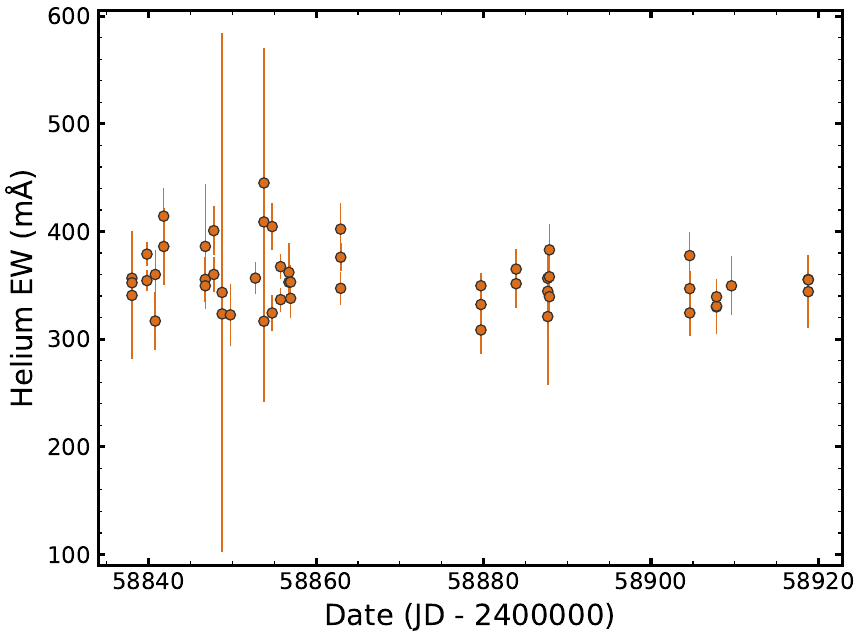}
    \caption{The \ewhe\ time series for the K2-100 intensive campaign. The data is fairly flat, and most of the variability shown is within the individual EW measurement uncertainties.}
    \label{fig:k2100_intensive}
\end{figure}

\textbf{K2-100:} K2-100 is an F6-dwarf Praesepe member that hosts a transiting planet \citep{mann2017}, and features the fastest rotation period in our sample besides the much younger V1298 Tau. However, it has \helium\ absorption strength and variability comparable to the rest of the ``older" targets in our sample. The intensive campaign for K2-100 comprises 24 visits (72 spectra) spanning 80 days, with a median visit separation of 1~day. There are two brief 16-day breaks in the observations in the middle of the campaign. This intensive campaign was taken to measure the mass of K2-100b, which will be presented in a future paper. Figure~\ref{fig:k2100_intensive} shows the intensive \ewhe\ time series of K2-100, which is fairly flat over the course of the observations, similarly to K2-136. We again generate mock time series assuming constant EW using each observation's measurement error to determine if the data variability is consistent with zero. We exclude from this calculation three data points with measurement errors in excess of 100~m\AA. The scatter for the data is $\sigma=23.9$~m\AA, which agrees with the simulated scatter assuming no intrinsic variability of $\sigma=26.0\pm4.6$~m\AA. Interestingly, there are no significant periodicities found for this data set. We may have expected rotational modulation due to its faster rotation period than K2-136, but this may be indicative of rotationally-smeared \helium\ absorption from the general chromospheric network or spatially distributed surface active regions.

For completeness, we compute the periodogram power spectra of all targets' full time series and see no significant periodicity, which is not surprising given their sparse cadence. In all, we only see any amount of rotational modulation in two targets: V1298 Tau with significant modulation over just one week and a half of the observations, and K2-136 with a periodigram power that is barely significant. While the \helium\ absorption strength variability seems to be dependent on the stellar age, and thus activity level and volatility, there is no significant relation between the \helium\ variability and stellar rotation. This may indicate that \helium\ line formation is not solely tied to spot or plage features, which are typically visualized as producing rotationally modulated stellar activity signals. Instead, \helium\ may be formed in active regions as well as from flares or other stochastic events in the broader chromospheric network, which essentially smears the \helium\ absorption changes across the stellar surface.

\section{Implications for the detection of young exospheres}\label{sec:exosphere_detection}

With the intrinsic stellar variability of the \helium\ triplet at young ages quantified, we can contemplate the possible effects of stellar activity on observations to find and characterize young helium exospheres. The stellar \helium\ triplet is demonstrably sensitive to activity, and intrinsic absorption changes could masquerade as an exosphere signal. The amplitude of the planetary \helium\ triplet absorption signal is set by the exosphere's metastable helium population and the atmospheric mass loss rate, both of which depend on the activity-sensitive high-energy radiative output of the host star \citep{oklopcic2019,poppenhaeger2022}. In this section, we discuss the feasibility of detecting young helium exospheres in the presence of stellar variability, and explore scenarios in which activity may affect exosphere observations.

\subsection{Relating equivalent width variability to line depth variability}\label{sec:ew_to_depth_var}

To assess the degree to which changes in the stellar \helium\ triplet may be confused with an exosphere's signal, we want to compare our sample’s intrinsic \helium\ triplet absorption strength variability to measured helium exosphere depths. However, we cannot directly compare our intrinsic \ewhe\ variability measurements to exosphere depths. Exosphere detections are most commonly presented as the peak in-transit excess absorption relative to an \oot\ reference spectrum at the center of the \helium\ triplet (typically the redder of the two resolved components). This would be equivalent to the variability in the {\it flux at the line center} rather than variability in the {\it integrated line strength}; our intrinsic \ewhe\ variability measures the latter.

For more accurate comparison we must convert the \ewhe\ variability to an equivalent relative absorption depth variability, which we call $\sigma_{D, \rm rel}$. This quantity also represents the excess absorption typically presented as an exosphere measurement. Let us consider a hypothetical spectral line with depth $D$ as a fraction of the continuum level. If there is a change in the depth of the line, called $\Delta D$, then the relative absorption depth variability can be calculated as:

\begin{equation}\label{eq:excess_absorption}
    \sigma_{D, \rm rel} = \frac{\Delta D}{1 - D}
\end{equation}

\noindent If we assume that only the line depth is changing (i.e. the line profile is the same), then any change in the equivalent width is proportional to the change in the line depth. We denote this as $\sigma_{\rm EW, rel}$, which is the relative change in the EW value:

\begin{equation}\label{eq:ew_variability}
    \sigma_{\rm EW, rel} = \frac{\Delta \rm EW}{\rm EW} = \frac{\Delta D}{D}
\end{equation}

\noindent We can measure $\sigma_{\rm EW, rel}$ by normalizing our measured intrinsic \ewhe\ variability by a target's typical \ewhe\ value. By combining Equations~\ref{eq:excess_absorption} and \ref{eq:ew_variability}, we can express the relative absorption depth variability (or excess absorption) in terms of the relative EW variability:

\begin{equation}\label{eq:corrected_ew_variability}
    \sigma_{D, \rm rel} = \frac{D}{1 - D} \sigma_{\rm EW, rel}
\end{equation}

With these equations, we can convert the measured intrinsic \ewhe\ variabilities for each target (given under column MAD$_{\rm intr}$[EW$_{\rm He}$] in Table~\ref{tab:helium} and described in Section~\ref{sec:variability}) into relative absorption depth variabilities that are directly comparable to exosphere measurements.

We normalize the intrinsic \ewhe\ variability by the typical \helium\ triplet absorption depth provided by the spectral fits from Section~\ref{sec:measure_ew}. The \ewhe\ we measure is of the full \helium\ triplet feature, which has two resolved components for 9 targets in our sample (the feature is unresolved for V1298 Tau due to its rapid rotation). For these 9 targets we use an ``equivalent single line" depth for the value $D$ in Equation~\ref{eq:corrected_ew_variability} by computing the depth required for an absorption line to have the same EW as the full feature EW. The width of the ``equivalent single line" is taken from the quadrature added variances of the individual \helium\ triplet components (using the median of the best fit values of the time series spectral fits). For V1298 Tau we use the depth of the unresolved line from the spectral fit.

We then substitute this intrinsic relative \ewhe\ variability into Equation~\ref{eq:corrected_ew_variability}. The resulting intrinsic relative absorption depth variabilities are listed in Table~\ref{tab:helium} as $\sigma_{D, \rm rel}$. Since we measure the intrinsic \ewhe\ variability using bootstrap sampling, we can calculate a distribution of intrinsic relative absorption depth variabilities for each target and estimate the uncertainty in expected variability. It is this distribution that is converted using Equation~\ref{eq:corrected_ew_variability}. The $\sigma_{D, \rm rel}$ values and uncertainties in Table~\ref{tab:helium} are the median and 16th-84th percentile confidence interval of each target's bootstrap distribution, representing the typical $1\sigma$ amplitude of the stellar variability signal and the uncertainty in our measurement of it. 

We also calculate a combined $\sigma_{D, \rm rel}$ value for the intermediate-aged stars in our sample (300~Myr to 1~Gyr) to best represent the variability at these ages by combining each target's individual $\sigma_{D, \rm rel}$ bootstrap distributions. These 8 targets all have similar \ewhe\ values and scatters which justifies the simple joining of their variability distributions. This combined distribution gives a $\sigma_{D, \rm rel}$ value of $0.65_{-0.65}^{+0.82}$, which agrees with a value of 0. This highlights that we cannot rule out the absence of intrinsic variability at these ages with our observations.

It is important to note that exosphere measurements typically only use the redder of the two resolved \helium\ triplet components. However, the triplet is unresolved for rapidly rotating stars such as V1298 Tau, which is why our \ewhe\ measurements use the full \helium\ triplet feature. To resolve this discrepancy, we recalculate $\sigma_{D, \rm rel}$ using only the red \helium\ triplet component for the 9 targets with a resolved feature. The EW calculation procedure from Section~\ref{sec:measure_ew} outputs the EW for the individual \helium\ triplet components. We then follow the steps in Section~\ref{sec:variability} to compute the red component's intrinsic variability, and substitute the red component's EW and variability into the steps listed previously in this section to calculate $\sigma_{D, \rm rel}$ in just the red component. For the depth value $D$ we use the best fit depth of the red component from Section~\ref{sec:measure_ew}'s spectral fits. The values for the red component are listed in Table~\ref{tab:helium} under the column labeled $\sigma_{D, \rm rel, red}$.

The $\sigma_{D, \rm rel}$ and $\sigma_{D, \rm rel, red}$ values agree within 1$\sigma$ for the 9 targets with resolved features. Thus, the relative intrinsic variability is not significantly different when considering the full feature or a subset of it. Any results concerning the \helium\ triplet should not be affected by the choice of which component(s) to measure. This is also encouraging for the ability to look for velocity-shifted signals with high-precision data. We note that $\sigma_{D, \rm rel, red}$ is typically lower precision and less constraining than $\sigma_{D, \rm rel}$. Using the full feature is ideal for variability measurements because it contains the most spectral information.

For the rest of the discussion we use the intrinsic relative absorption depth variabilities calculated with the full \helium\ triplet feature ($\sigma_{D, \rm rel}$ in Table~\ref{tab:helium}). We make this decision because V1298 Tau has an unresolved feature and is crucial to our analysis as the youngest star in the sample. This also keeps consistent our use of the full \ewhe\ across the paper. The agreement between the values calculated with the full feature and only the red component further justifies this decision.

\subsection{Comparisons with known helium exospheres}

\begin{figure}
    \centering
    \includegraphics[width=\columnwidth]{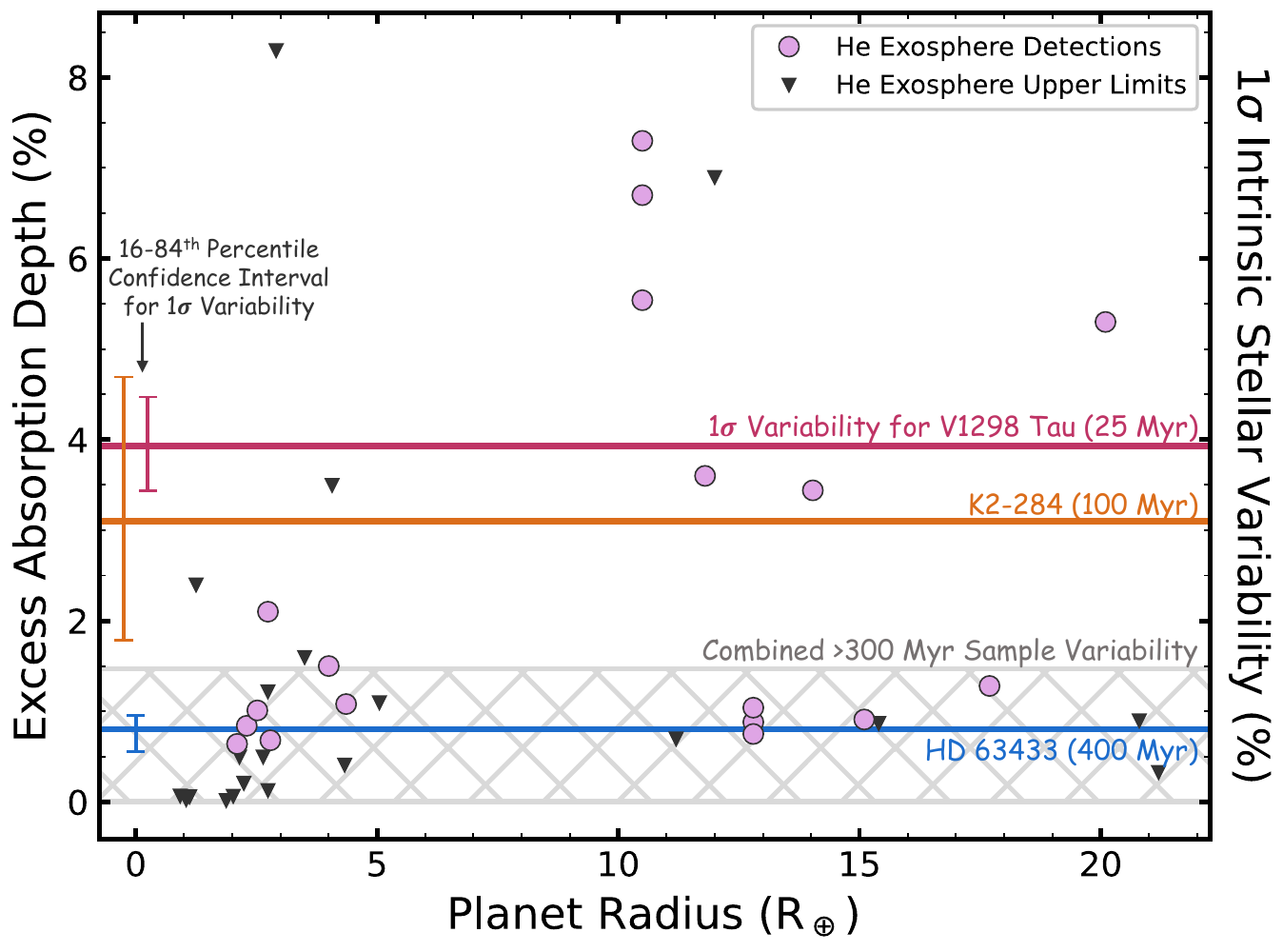}
    \caption{Literature helium exosphere detections (circles) and upper limits (triangles) compared to our measurements of the 1$\sigma$ intrinsic stellar variability level (see Section~\ref{sec:ew_to_depth_var} for a description of the calculation). To show the age dependence, we plot the variability of V1298 Tau, K2-284, and HD 63433 as separate horizontal lines. The error bars at the left end of the plot show the 16th-84th percentile confidence interval on our measurement of the 1$\sigma$ intrinsic stellar variability. The gray hatched region shows the 16th-84th percentile confidence interval on the variability measured with the combined $>300$~Myr sample, which agrees with no intrinsic variability. All but a few exosphere detections are smaller than variability at the youngest ages, but by 400 Myr all exosphere detections are comparable to or larger than the intrinsic stellar variability. Stellar variability should not preclude detection at these intermediate ages. The literature data is presented in Table~\ref{tab:lit_he_exospheres}.}
    \label{fig:exosphere_comp}
\end{figure}

Figure~\ref{fig:exosphere_comp} shows helium exosphere detections and upper limits from the literature as percentage excess absorption depths, plotted alongside some of our measurements of the intrinsic relative absorption depth variability ($\sigma_{D, \rm rel}$) for our sample plotted as horizontal lines. The $\sigma_{D, \rm rel}$ quantity represents a 1$\sigma$ variability amplitude and thus it is not directly comparable to a detection limit, although it could be scaled to serve as one if an assumption is made for the underlying depth variability distribution.

We have separately plotted 4 different $\sigma_{D, \rm rel}$ values from our sample to visualize the age dependence. We show the $\sigma_{D, \rm rel}$ value (solid line) and its 16th-84th percentile confidence interval (error bar at plot's left side) for V1298 Tau to represent the youngest and most active stars, HD 63433 to represent ages when variability approaches the field’s, and K2-284 to represent ages between these two extremes. To compare to the upper limit on intrinsic variability calculated for stars older than 300 Myr, we show the 16th-84th percentile confidence interval of the combined intermediate age $\sigma_{D, \rm rel}$ as the gray hatched region. Note that this range agrees with 0, meaning no intrinsic variability to prevent exosphere detection.

A vast majority of all currently detected helium exospheres have depths significantly smaller than the intrinsic variability measured for stars with $\tau \lesssim 120$~Myr (V1298 Tau and K2-284). There are however a handful of extreme detections, with depths $\gtrsim4\%$, that are comparable in amplitude to or even larger than the variability at these youngest ages. From this, we conclude that comparing spectra from across observing seasons (even separated by days) is inadequate for detecting helium exospheres at ages below $\sim120$~Myr, although large transit absorption signals like WASP-107 b's can likely be ruled out \citep{allart2019,kirk2020,spake2021}. Directly contemporaneous \oot\ comparison observations and the use of multiple transit detections are likely needed for significant detections of the youngest exospheres.

We show HD 63433 separately from the rest of the sample at ages older than 300~Myr because it is the most precise variability measurement that we make. All exosphere detections are either larger than or in agreement with the 16th-84th percentile confidence interval of HD 63433's variability, and only the most precise upper limits are smaller. This is promising for the detection of helium exospheres at intermediate ages and agrees with our conclusion that stars of this age exhibit similar \helium\ triplet behavior as the field. HD 63433 may not be representative of all stars in this age range, so we also show the 16th-84th percentile confidence interval of the intrinsic variability for the ensemble of stars older than 300~Myr. While the top of this interval is larger than a majority of the known helium exosphere depths, the measurement from this stellar ensemble agrees with a total lack of intrinsic stellar variability that would ensure the detectability of all helium exospheres (barring significant instrumental or measurement errors).

It is very important to note that the $\sigma_{D, \rm rel}$ values we calculate are essentially a 1$\sigma$ amplitude for the intrinsic relative depth variability. This quantity does not directly provide an exosphere detection limit because a statistically significant measurement typically requires precision $>1\sigma$. This is complicated by the fact that while an exosphere may or may not be present, a transit is not a random event. It is difficult to determine a detection limit from these values, so the comparison with literature exospheres is meant to be illustrative. One suggestion is to adopt $1.7\sigma_{D, \rm rel}$ as a limit, which represents an estimated one-sided 95th percentile limit on the relative depth change, if the variability follows a Gaussian distribution.

There are caveats to discuss that both positively and negatively impact the potential detection of young helium exospheres. We so far discuss the \textit{long-term, time-averaged} variability, which represents the typical difference in EW from the median value averaged across long observation baselines. Crucially, we show in Section~\ref{sec:timescales} that \ewhe\ variability on the shortest timescales ($\Delta t\la30$~min) is statistically consistent with no intrinsic variability across age. \Oot\ observations directly surrounding transit should further decrease the prospect of activity-driven confusion in interpreting results. There are also reasons why young exosphere signals may be larger than their older counterparts. Young stars have greater high-energy radiative output, which should increase both the exosphere's metastable helium population \citep[leading to more \helium\ triplet absorption;][]{oklopcic2019,poppenhaeger2022} and atmospheric mass loss rate \citep{lopez2012,lopez2013,owen2017}.

However, many transits are comparable in duration to or longer than the timescale of increased variability. If the intrinsic stellar feature is changing over the duration of a transit, the \oot\ reference spectra may not be truthful representations of the base stellar absorption, and any exosphere interpretation will be affected. There may also be spurious signals on these short timescales caused by observations with inopportune timing, such as from a coincident activity event (e.g. flares). We discuss such scenarios further in Section~\ref{sec:variability_scenarios} below. We also note that our assumption of a stable line profile to generate Equation~\ref{eq:corrected_ew_variability} is not wholly accurate, so the comparison made in Figure~\ref{fig:exosphere_comp} is not exact.

Overall, detecting ``intermediate-aged" helium exospheres ($\tau \gtrsim 300$~Myr) is not necessarily precluded by stellar activity, even with observations spanning an observing season. However, potential confusion is minimized with well-timed \oot\ reference observations and by confirming the exosphere absorption in multiple transits. Stellar variability likely would preclude detection at the youngest ages ($\tau\la100$~Myr) without carefully planned observations and significant exosphere absorption.

\subsection{Scenarios where activity may still affect exosphere observations}\label{sec:variability_scenarios}

There are still situations in which stellar activity-driven variability in the \helium\ triplet could complicate exosphere detection. In particular, we are concerned with scenarios where the timescales of stellar variability and exosphere observations overlap. While our measurement of \ewhe\ scatter at the very shortest timescales is consistent with no intrinsic variability, it is important to note that there are still situations that may lead to actual intrinsic variability in \ewhe\ on these timescales. Below, we discuss four scenarios that may affect the detection of young exospheres, although this list is hardly exhaustive and meant to be emblematic of potential issues with the \helium\ triplet.

\textbf{1) Flares:} Stellar \helium\ triplet absorption is enhanced by flares \citep{andretta2008,sanzforcada2008,kobanov2018}. If a flare occurs near transit, there may be an increase in the \helium\ triplet absorption strength that could be interpreted as an exosphere, despite being stellar in nature. This would be most likely for very young stars that feature the greatest flare rates. In our sample this includes V1298 Tau, for which \citet{vissapragada2021} observed a flare during the transit of one of its planets, and could not distinguish between the flare or an exosphere as the cause for increased absorption. Flares could also increase the atmospheric mass loss rate \citep{wang2021}, leading to an increase in the exosphere \helium\ triplet absorption lagged from the flare by a few hours. In that case, enhanced absorption would be a result of \textit{both} the star and the exosphere. It may be possible to disentangle these signals given the lifetime of each, but the temporal response of the stellar \helium\ triplet feature to a flare is unknown.

\textbf{2) Lack of rotational coherence:} We find nearly no significant signals at the stellar rotation period in our data, implying that changes in the \helium\ triplet absorption are not rotationally coherent (particularly for longer than a few rotations). This could be detrimental to exosphere detection, as the activity-induced variability could not be modeled at the known rotation period and subtracted. This also means that the \helium\ triplet could not be Doppler mapped to identify distinct causative active regions. Other traditional spectral activity indicators that do correlate with rotation may not track changes in the \helium\ triplet, and thus could not be used in models of activity's effect on the \helium\ triplet.

\textbf{3) Longer baseline \oot\ comparison:} \Oot\ comparison spectra may simply need to be taken further separated from mid-transit than the $\sim5$~hr timescale of increased intrinsic stellar \helium\ absorption variability. One reason could be a transit that is on the order of or longer than this timescale. The median transit duration of the known young transiting planets is 3.2~hr and the longest duration is 7.5~hr. Extended \oot\ baseline could be long enough from parts of the transit when stellar variability may introduce changes in the \helium\ absorption strength. Depending on the transit timing, object visibility, and observing conditions, it might be necessary to obtain \oot\ comparison spectra on a different night, which is more susceptible to activity contamination. This also would affect comparing absorption depth changes across multiple transits.

\textbf{4) Exosphere structure:} An extended exosphere may affect its own detection, particularly for a tail-like structure that effectively increases the helium transit duration compared to the white-light transit. Tails created by atmospheric mass loss have long been predicted \citep{schneider1998}, and have now been observed in helium gas for multiple planets \citep{nortmann2018,alonsofloriano2019,spake2021}. Bow shocks and up-orbit accretion flows could also produce a ``leading tail" \citep{matsakos2015}, which has been observed for one helium exosphere \citep{czesla2022}. By lengthening the transit, these extended exospheres increase the likelihood of activity-induced variability occurring during transit observations, as explained in point 3 above. Such a scenario would complicate the definition of ``\oot" observations, and necessitate more baseline observations further from transit. One can imagine an extreme situation of a compact multi-planet system where tails from multiple planets could lead to overlapping extended transits \citep[e.g. TRAPPIST-1, although no helium exosphere was detected for 3 of its planets;][]{krishnamurthy2021}. This would result in a ``constant" presence of helium exosphere material in the system. While this situation is not necessarily common, such multi-planet systems (like V1298 Tau) are being observed, so it is plausible enough to encourage care when considering the timing of transit observations.

\subsection{Companion observations for planning campaigns and mitigating He variability}

A reconnaissance NIR spectrum to measure the \ewhe\ of planet hosts would help to prioritize targets for exosphere observing campaigns. High \ewhe\ itself, which is enhanced by stellar activity, could indicate a higher likelihood of a detectable helium exosphere. Activity is accompanied by greater stellar high-energy radiative output, which would excite more metastable helium in the exosphere to produce a deeper planetary absorption signal. \ewhe\ is less informative for active stars with $\tau \gtrsim 300$~Myr, though, because the absorption strength saturates at intermediate ages. The \ewhe\ reconnaissance may be more useful for older, less active stars where the PR mechanism (and thus coronal radiation) will dominate the metastable helium excitation. It is important to note that activity in the youngest stars \textit{also} indicates increased stellar variability, which would impede exosphere observations. Regardless, stronger \helium\ absorption of a host star may be useful as an indicator for the likelihood of a detectable helium exosphere.

There are also companion observations that could help to mitigate contamination of exosphere observations from stellar activity-driven variability. Extended \oot\ monitoring would establish a more reliable baseline stellar \helium\ absorption strength, and could be used to identify the timescale and amplitude of intrinsic stellar variability. Long-term monitoring could also be used to distinguish between periods of increased and decreased variability for planning observations (as we see in Section~\ref{sec:intensive_campaigns}). Ideally, one would observe consecutive transits with baseline observations taken before the first transit through after the last transit. This would provide a long baseline for comparison, as well as multiple transits to increase the reliability of any detection. Simultaneous photometry could be used to monitor for strong flares that may affect the stellar absorption near transit (leading to intrinsic short-timescale variability), as was done for an observation of V1298 Tau by \citet{vissapragada2021}. 

Beyond this, further studies of the stellar \helium\ triplet are needed to connect the feature to established activity proxies that may help de-trend the intrinsic stellar variability in the line. A larger sample of stars at younger ages (particularly between 50 and 300 Myr) would further establish the behavior of the \helium\ triplet in youth and the feasibility of exosphere detections.

\section{Conclusions and future work}\label{sec:conclusions}

Young planets hold unique potential to better our understanding of the formation and evolution of exoplanet atmospheres. Since its mission began, TESS has followed K2 in discovering young planetary systems around bright stars that are well-suited for atmospheric characterization. Critically, some of these systems have been found in new stellar associations with ages between 100 and 400 Myr \citep[e.g.][]{newton2021,tofflemire2021,hedges2021,dong2022,newton2022}. With the continued use of precision NIR spectrographs, the \helium\ triplet will persist as a crucial probe of the mass loss from these young planets. These follow-up observations will require knowledge of the intrinsic stellar variability of the \helium\ triplet because the host stars have high activity levels.

In this paper, we present an initial study of the \helium\ triplet in youth using unique NIR spectroscopic data from HPF for a subset of the known young transiting planet hosts. We measured the \helium\ triplet absorption strength and variability of our sample to characterize the feature's relationship with stellar activity, and discussed implications for exosphere detections. To summarize our results:

\begin{enumerate}
    \item We developed a self-calibration technique to subtract strong telluric sky emission lines from HPF spectra, which are not adequately subtracted using just the sky fiber spectra. This is important as there is a very strong sky line that can overlap the \helium\ triplet, making it harder to measure the feature's absorption strength.
    
    \item We find that the \ewhe\ is enhanced for young stars relative to inactive dwarfs. There is no effective temperature dependence, but there is a relation with stellar rotation period: there is stronger absorption at shorter rotation periods. While all of our stars are young, activity levels are higher at the youngest ages and most rapid rotation. Despite a likely significant difference in the high-energy radiative output across our sample, we find a plateau in \ewhe\ from $P_{\rm rot} \sim 5-20$~days. We conclude that in young and active chromospheres, metastable helium is populated primarily through collisional excitation, and that contributions from the PR mechanism only dominates at the high and low relative activity regimes.

    \item The intrinsic stellar variability in \ewhe\ is large at young ages ($\tau \lesssim 100$~Myr), and decreases to a plateau comparable to the expected field variability at $\tau \gtrsim 300$~Myr that is in agreement with a lack of intrinsic variability. While the intrinsic variability of the 400~Myr HD 63433 is measurable at $\sim2$\% to a precision of $\sim0.5$\%, the measurements for 6 of the other 7 ``older" stars lack \ewhe\ precision to rule out no intrinsic variability. This variability is caused by volatility and evolution of the stellar activity level, such as by flares or changing surface heterogeneities, which is most common at the youngest ages.
    
    \item The stellar \ewhe\ variability is smallest on the shortest timescales (within 30 min), and is consistent with no intrinsic variability on this timescale considering all data. This gives us confidence in the ability to detect young exospheres with immediate \oot\ comparison observations. However, the variability increases within just 5 hours for the youngest star in our sample (V1298 Tau) and within a day for the rest of the sample, perhaps from the coherence of flare-induced absorption changes. Comparing spectra of young stars from night-to-night may introduce stellar variability that can mask or masquerade as an exosphere signal. The convolution of stellar and planetary signals must be considered when longer baseline \oot\ observations are taken, particularly if the exosphere has extended structure, such as a tail.
    
    \item We find little evidence for periodicity with the stellar rotation in \ewhe\, implying that the chromospheric regions in which the line forms are not rotationally localized. We do, however, see stellar rotation periodicity within the first week and a half of a month-long intensive campaign of V1298 Tau, featuring extreme absorption strength changes. The \helium\ triplet is most variable and complicated at the youngest ages, and may have rotational modulation during extreme flaring events or in the presence of large, persistent high contrast surface heterogeneities.

    \item Youth does not necessarily preclude the detection of helium exospheres, except at the youngest ages. The intrinsic variability at ages above 300~Myr is smaller than or comparable to all current helium exosphere detections, but for ages below 120~Myr the intrinsic variability is larger than all but the most extreme helium exosphere detections. Variability also further decreases at the shortest timescales, improving detectability. However, there may still be confusion between enhanced absorption from exospheres and stellar active regions or flares. Care must be taken when searching for exosphere signals when observations are separated by longer than a few hours, as is often the case for a transit, because variability can be significant on timescales greater than a day. Continuous spectroscopic and photometric monitoring around transit would help to mitigate the deleterious effect of stellar activity.

\end{enumerate}

Our understanding of the \helium\ triplet of young and active stars is still incomplete. More targets must be observed and uniformly analyzed to cover our sample's gaps in age, spectral type, rotation period, and activity level. Mapping \ewhe\ at more densely sampled stellar parameters, particularly rotation period, would further elucidate the origin of the line in young chromospheres. Only two of our targets are young enough to feature enhanced \helium\ triplet variability. More young targets ($\tau \la 300$~Myr) are needed to robustly assess the feasibility of detecting young exospheres and determine variability timescales that help to best plan observing campaigns. Additionally, intensive campaigns of a select few highly active stars could be used to search for activity cycles and rotational coherence in \ewhe.

Other observations that are contemporaneous with NIR spectroscopy would complement the conclusions we draw about the \helium\ triplet in youth. Photometric monitoring could assist in connecting \helium\ triplet variability to spot modulation and evolution, and search for rotational or long-term cyclic modulation in the \helium\ triplet. X-ray or UV observations would provide instantaneous measurements of the high-energy coronal radiation that may drive the excitation of the helium metastable state, and definitively distinguish contributions from the PR and CE mechanisms. Visible spectroscopy would provide measurements of established chromospheric activity indicators (such as Ca~\textsc{ii} H and K, and H-$\alpha$), mapping lines that are formed at varying chromospheric heights and tying helium variability to well-studied probes of activity.

NIR precision RV measurements are still relatively new, and there are no well-established activity indicators in the NIR bandpass that are analogous to Ca~\textsc{ii} H and K in the visible. The lack of rotational coherence in our observations of the \helium\ triplet calls into question whether or not it would work well as an activity indicator to de-trend RV jitter. It may provide information about second-order activity effects on the stellar spectral line profile, but may not be able to act alone as a NIR activity indicator for precision RV studies. An in-depth analysis connecting \ewhe\ to RVs for our sample is beyond the scope of this paper, but will be featured in a future work.

Time series studies of young exoplanet hosts provide crucial data for understanding young stellar structure, planet formation and evolution, and the effects of stellar activity on exoplanet observations. In particular, the recent developments in NIR precision spectroscopy have opened an entirely new window into studies of stars and exoplanets, but the results of these ongoing programs are nascent \citep{reiners2018,donati2020,tran2021}. With advancements in panchromatic observing capabilities, it is ever more pressing to expand the frontiers of stellar astrophysics to best understand the manifestation of stellar activity in extreme precision exoplanet observations.

\begin{acknowledgments}
We thank the anonymous referee for a thorough review that improved the manuscript and strengthened our conclusions. We thank Michael Gully-Santiago, Antonija Oklop\v{c}i\'{c}, Aaron Rizzuto, Quang Tran, and Chad Bender for useful discussions regarding the \heliumAA\ triplet feature, telluric correction, best practices for using HPF data, and clarity of the writing. We thank the HPF team for providing the high-quality data products necessary for the work presented in the manuscript. We are also very grateful for the resident astronomers and telescope operators at the HET, particularly Steven Janowiecki, for help in planning HPF observations and using our time allocation to its maximum potential.

These results are based on observations obtained with the Habitable-zone Planet Finder Spectrograph on the HET. The Hobby–Eberly Telescope is a joint project of the University of Texas at Austin, the Pennsylvania State University, Ludwig-Maximilians-Universität München, and Georg-August Universität Gottingen. The HET is named in honor of its principal benefactors, William P. Hobby and Robert E. Eberly. The HET collaboration acknowledges the support and resources from the Texas Advanced Computing Center. We would like to acknowledge that the HET is built on Indigenous land. Moreover, we would like to acknowledge and pay our respects to the Carrizo \& Comecrudo, Coahuiltecan, Caddo, Tonkawa, Comanche, Lipan Apache, Alabama-Coushatta, Kickapoo, Tigua Pueblo, and all the American Indian and Indigenous Peoples and communities who have been or have become a part of these lands and territories in Texas, here on Turtle Island.

This research has made use of the SIMBAD database, operated at CDS, Strasbourg, France, and NASA’s Astrophysics Data System Bibliographic Services.

\end{acknowledgments}

\vspace{5mm}
\facility{HET (HPF)}

\software{Astropy \citep{astropy2013, astropy2018}, barycorrpy \citep{kanodia2018}, ipython \citep{ipython}, jupyter \citep{jupyter}, matplotlib \citep{matplotlib}, NumPy \citep{numpy}, pandas \citep{pandas}, SciPy \citep{scipy}, TelFit \citep{gullikson2014}}

\appendix
\restartappendixnumbering

\section{Literature helium exosphere observations}\label{app:lit_helium}

In Table~\ref{tab:lit_he_exospheres} we show the literature helium exosphere observation data points plotted in Figure~\ref{fig:exosphere_comp}. These data come from studies using high resolution spectroscopy, and include detections and upper limits. If there are multiple published observations for a planet, whether detections or upper limits, we include all of them in the table and the corresponding figure.

\startlongtable
\begin{deluxetable*}{ccccc}
\tablecaption{Literature helium exosphere observations\label{tab:lit_he_exospheres}}
\tablehead{
\colhead{Planet} & \colhead{Excess Absorption \%} & \colhead{Instrument} & \colhead{Limit?\tablenotemark{a}} & \colhead{Reference}
}
\startdata
HD 189733 b & 0.88 & CARMENES & & \citet{salz2018}\\
HD 189733 b & 1.04 & CARMENES & & \citet{nortmann2018}\\
HD 189733 b & 0.75 & GIANO-B & & \citet{guilluy2020}\\
HAT-P-11 b & 1.08 & CARMENES & & \citet{allart2018}\\
WASP-69 b & 3.6 & CARMENES & & \citet{nortmann2018}\\
HD 209458 b & 0.91 & CARMENES & & \citet{alonsofloriano2019}\\
WASP-107 b & 5.54 & CARMENES & & \citet{allart2019}\\
WASP-107 b & 7.3 & NIRSPEC & & \citet{kirk2020}\\
WASP-107 b & 6.7 & NIRSPEC & & \citet{spake2021}\\
GJ 3470 b & 1.5 & CARMENES & & \citet{palle2020}\\
HAT-P-32 b & 5.3 & CARMENES & & \citet{czesla2022}\\
HD 73583 b & 0.68 & NIRSPEC & & \citet{zhang2022a}\\
GJ 1214 b & 2.1 & CARMENES & & \citet{orellmiquel2022}\\
GJ 1214 b & 0.13 & NIRSPEC & UL & \citet{kasper2020}\\
GJ 1214 b & 1.22 & NIRSPEC & UL & \citet{spake2022}\\
WASP-52 b & 3.44 & NIRSPEC & & \citet{kirk2022}\\
WASP-177 b & 1.28 & NIRSPEC & & \citet{kirk2022}\\
TOI 1430.01 & 0.64 & NIRSPEC & & \citet{zhang2022d}\\
TOI 1683.01 & 0.84 & NIRSPEC & & \citet{zhang2022d}\\
TOI 2076 b & 1.01 & NIRSPEC & & \citet{zhang2022d}\\
TOI 2076 b & 1 & IRD & UL & \citet{gaidos2023}\\
GJ 436 b & 0.41 & CARMENES & UL & \citet{nortmann2018}\\
KELT-9 b & 0.33 & CARMENES & UL & \citet{nortmann2018}\\
K2-100 b & 1.6 & IRD & UL & \citet{gaidos2020a}\\
WASP-127 b & 0.87 & Gemini/Phoenix & UL & \citet{dossantos2020}\\
AU Mic b & 3.5 & IRD & UL & \citet{hirano2020}\\
TOI 1728 b & 1.1 & HPF & UL & \citet{kanodia2020}\\
GJ 9827 b & 0.067 & NIRSPEC & UL & \citet{kasper2020}\\
HD 97658 b & 0.21 & NIRSPEC & UL & \citet{kasper2020}\\
55 Cnc e & 0.02 & NIRSPEC & UL & \citet{zhang2021}\\
TRAPPIST-1 b & 0.06 & IRD & UL & \citet{krishnamurthy2021}\\
TRAPPIST-1 e & 0.07 & IRD & UL & \citet{krishnamurthy2021}\\
TRAPPIST-1 f & 0.03 & IRD & UL & \citet{krishnamurthy2021}\\
WASP-76 b & 0.9 & CARMENES & UL & \citet{casasayas2021}\\
K2-136 c & 8.3 & IRD & UL & \citet{gaidos2021}\\
WASP-80 b & 0.7 & GIANO-B & UL & \citet{fossati2022}\\
HD 63433 b & 0.5 & NIRSPEC & UL & \citet{zhang2022b}\\
HD 63433 c & 0.5 & NIRSPEC & UL & \citet{zhang2022b}\\
TOI 3757 b & 6.9 & HPF & UL & \citet{kanodia2022b}\\
TOI 1807 b & 2.4 & IRD & UL & \citet{gaidos2023}\\
\enddata
\tablenotetext{}{The exosphere measurements are mostly ordered by publication date, and separated by detection vs. upper limit. The exceptions are when there are multiple measurements for a single planet; in that case, all measurements are grouped together (regardless of date or distinction between detection and upper limit). Only high resolution spectroscopic observations are included.}
\tablenotetext{a}{Marked as ``UL" if the measurement is an upper limit on the excess absorption.}
\end{deluxetable*}

\bibliography{planets_hpf}

\end{document}